\documentclass{rsproca}
\begin{document}
\title{Stationary localized structures and the effect of the delayed feedback in the Brusselator model}
\author{B. Kostet$^{1}$, M. Tlidi$^{1}$, F. Tabbert$^{2,3}$, T. Frohoff-H\"ulsmann$^{2,3}$, S. V. Gurevich$^{2,3}$,E. Averlant$^{1}$, R. Rojas$^{4}$, G. Sonnino$^{1}$, K. Panajotov$^{5,6}$}
\address{Universit\'e libre de Bruxelles (U.L.B.), Facult\'e des Sciences, Campus Plaine, B-1050 Bruxelles, Belgium $^{1}$\\
Institute for Theoretical Physics, University of M\"unster, Wilhelm-Klemm-Str. 9, D-48149 M\"unster, Germany$^{2}$\\
Center for Nonlinear Science (CeNoS), University of M\"unster, Corrensstr. 2, D-48149 M\"unster, Germany$^{3}$\\
Intituto de F\'isica, Pontificia Universidad Cat\'olica de Valpara\'iso,
casilla 4059, Valpara\'iso, Chile$^{4}$\\
Vrije Universiteit Brussel, Department of Applied Physics and Photonics, Brussels Photonics (B-PHOT), Pleinlaan 2, B-1050 Brussels, Belgium $^{5}$ \\
Institute of Solid State Physics, 72 Tzarigradsko Chaussee Boulevard, 1784 Sofia, Bulgaria$^{6}$
}
\subject{42.55.Px, 42.60.Mi, 05.45.-a,02.30.Ks}
\corres{G. Sonnino\\\email{gsonnino@ulb.ac.be}}
\begin{abstract}
 The Brusselator reaction-diffusion model is a paradigm for the understanding of dissipative structures in systems out of equilibrium. In the first part of this paper, we investigate the formation of stationary localized structures in the Brusselator model. By using numerical continuation methods in two spatial dimensions, we establish a bifurcation diagram showing the emergence of localized spots. We characterize the transition from a single spot to an extended pattern in the form of squares. In the second part, we incorporate delayed feedback control and show that delayed feedback can induce a spontaneous motion of both localized and periodic dissipative structures. We characterize this motion by estimating the threshold and the velocity of the moving dissipative structures.
\end{abstract}

\maketitle
\section{Introduction}
The spontaneous emergence of dissipative structures in chemical systems arises from a principle of self-organization that can be either in space and/or in time  \cite{Balesu,Lefever,Prigogine}. I. Prigogine demonstrated that chemical oscillations are  perfectly compatible with the second principle of thermodynamics extended to far from equilibrium systems (see overviews on this issue  \cite{Lefever_PTRSA,ERneux_PTRS}). Spatial dissipative structures generally evolve on macroscopic scales and can only be maintained by continuous application of a non-equilibrium constraint. A classic example of spatial self-organization has been provided by A. Turing \cite{Turing}. This mathematical prediction has been supported by I. Prigogine's explanation of the physical mechanisms underlying the formation of dissipative structures.  The Turing-Prigogine dissipative structures are characterized by an intrinsic wavelength. The formation of spatial structures is attributed to the balance between a nonlinear process originating from a chemical reaction that tends to amplify spatial fluctuations and a diffusion process that tends, on the contrary, to restore uniformity. When a system operates far from equilibrium, dissipation plays an important role in the stabilization of self-organized structures. Dissipative structures in chemical systems  have been experimentally observed  thanks to the development of chemical open reactors with a Chlorite-Iodide-Malonic-Acid reaction \cite{DeKepper,Swinney}. The above  mechanism has been widely used to explain the formation of stationary periodic dissipative structures not only in chemistry but in various biological and physical systems~\cite{Rev5,Rev6,Rev9,Rev11,Rev12,Rev13,Rev16,Lugiatobook,Tlidi-Krassi,Rev8,Deneubourg,ga96}.

Besides the spatially periodic structures, the same mechanism predicts the possible existence of aperiodic and localized structures. The latter consists of isolated or randomly distributed spots surrounded by regions of the uniform state \cite{kURAMOTO-kOGA,Malomed_Ne,Tlidi_94,Dewel_94,Hilali}. They correspond to the confinement of energy, chemical concentration, light or biomass density in one or more spatial dimensions (see recent overviews on this issue \cite{Leblond-Mihalache,Tlidi-PTRA,Lugiatobook,Rev16}). A vast amount of theoretical work has been carried out on localized structures in chemical reaction-diffusion systems, \cite{RD_LS_1,RD_LS_2,RD_LS_3,RD_LS_4,RD_LS_5,RD_LS_6,RD_LS_7,TVL_11,Lier2013,Tlidi_Entropy,RD_LS_8} while there is still a lack of experimental evidence of stable localized spots. This is because isolated localized structures (LSs) often exhibit a self-replication phenomenon that affects the circular shape of the LS \cite{Pearson,LEE} and leads to the formation of an extended patterned state.

In this paper, we  consider the paradigmatic Brusselator reaction-diffusion model. We report on stationary two-dimensional solutions of the model consisting of localized spots. We provide a bifurcation analysis of the localized solutions, which arise in a subcritical pitchfork bifurcation and undergo a self-replication instability leading to bound states of localized solutions. We use two-dimensional continuation techniques provided by the matlab package pde2path \cite{pde2path} to construct the bifurcation diagram associated with two-dimensional localized structures. In the last part of the manuscript, we investigate the effect of delayed feedback on the stability of localized structures.  We show that both periodic and localized structures exhibit a spontaneous motion in an arbitrary direction, whereas in the absence of time-delayed feedback, a single spatial spot and periodic patterns remain stationary. The bifurcation diagram associated with a two-dimensional single localized spot is constructed. We characterize the motion by estimating the speed of moving dissipative structures and by analyzing the threshold associated with the onset of motion as a function of the feedback parameters.

The paper is organized as follows: after the introduction, we analyze the bifurcation structure of  localized spots in section 2. In section 3, we implement delayed feedback control. We analyze the impact of delayed feedback on the linear regime. In section 4, we characterize the transition from a stationary to a moving localized spot. In section 5, we conclude.



\section{Localized structures in the Brusselator model}
The Brusselator model is a classical reaction-diffusion system, proposed by I. Prigogine and R. Lefever~\cite{Lefever}.  The evolution equations of the Brusselator model read
\begin{eqnarray}
\frac{\partial X}{\partial t} &= &\frac{\partial X^{2} }{\partial
x^{2}} +\frac{\partial X^{2} }{\partial
y^{2}}+ A-(B+1)X+X^2Y \nonumber\\
\frac{\partial Y}{\partial t} &= &D\left(\frac{\partial Y^{2} }{\partial x^{2}} +\frac{\partial Y^{2} }{\partial y^{2}}\right) + BX -X^2Y. \label{eq:Bru}
\end{eqnarray}
Here, $X=X(x,\,y)$ and $Y=Y(x,\,y)$ denote the concentrations of the interacting chemical species, whereas $A$ and $B$ are externally controlled concentrations which play the role of control parameters. The parameter $D$ is the ratio between the diffusion coefficients  $D_y$ and $D_x$ of the $Y$ and $X$ concentrations, respectively. 

\begin{figure}[bbp]
	\begin{center}
		\includegraphics[width=1.0\textwidth]{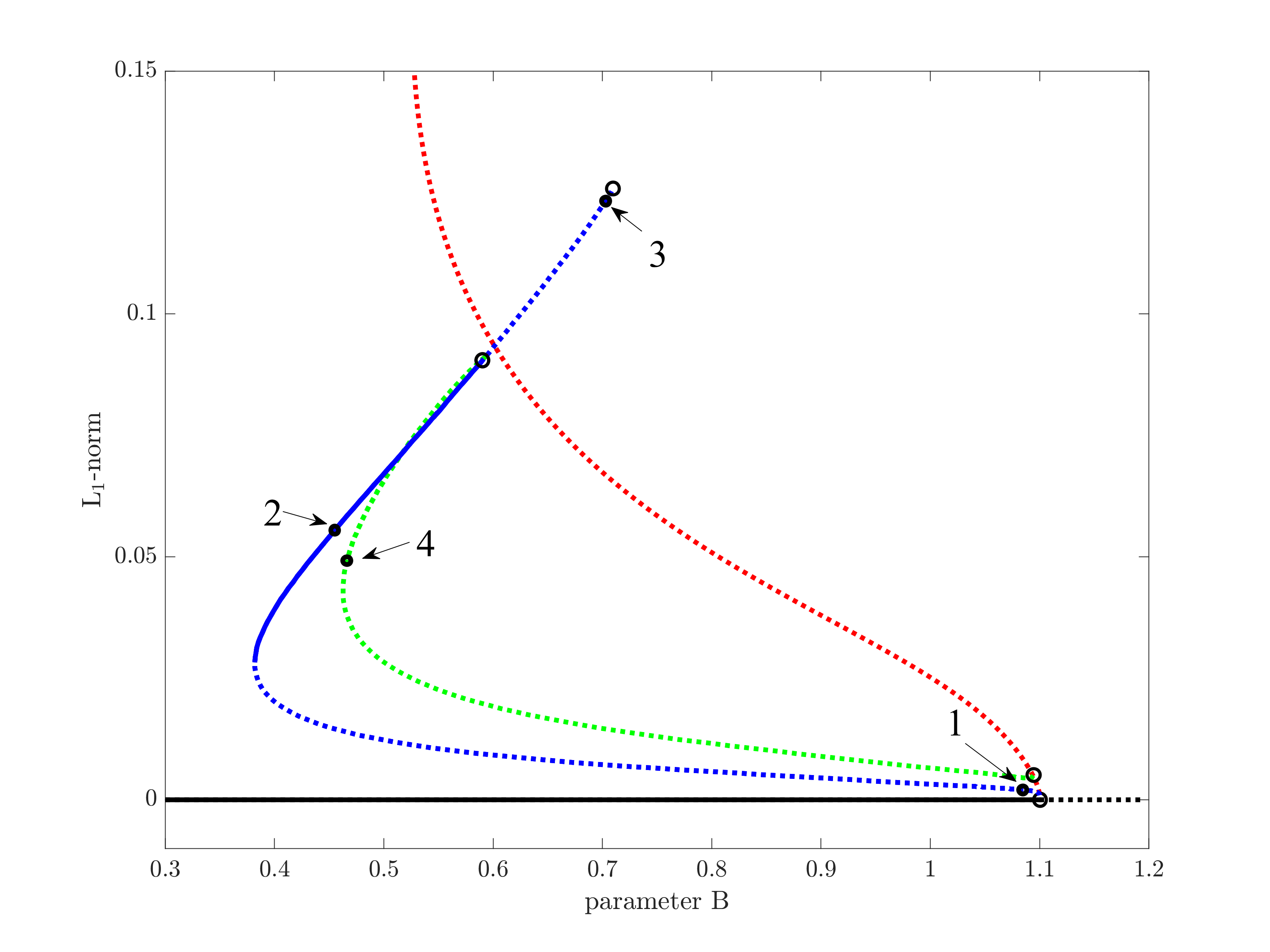}
	\end{center}
	\caption{{Bifurcation diagram for a single localized solution of the two-dimensional Brusselator model \eqref{eq:Brudelay} without time-delayed feedback obtained by numerical continuation. Depicted is the $L_1$-norm with respect to the mean field of the first variable $X$ for different solutions. The continuation parameter is $B$. Fixed parameters are the domain size $L_x=L_y=80$, and $D=150$, $A=0.6$. The black line corresponds to the homogeneous solution. At the Turing-Prigogine point periodic solutions bifurcate from the homogeneous solution. The dotted red line represents the periodic rhombic solution which bifurcates subcritically. Shortly after, the single localized solution (dotted blue line) bifurcates from the periodic rhombic solution in a supercritical pitchfork bifurcation. The localized solution becomes stable in a fold (solid blue line) and then again unstable in a subcritical pitchfork bifurcation, in which a ringlike (dotted blue line) and a self-replicated solution (dotted green line) branch off. Solution profiles for the marked positions can be found in Fig. \ref{bifsol}.}}
\label{bifdiag}
\end{figure}
The homogeneous steady state of \eqref{eq:Bru} is $(X_s,\,Y_s)=(A,\, B/A)$. It can  undergo a Turing-Prigogine bifurcation leading to the formation of a stationary oscillation in space or an Andronov-Hopf instability leading to time-periodic chemical oscillations. The threshold of the Turing-Prigogine bifurcation is given by $B=(1+A/\sqrt{D})^2$, $D^{-1}<1$ whereas the Andronov-Hopf instability occurs if  $B>1+A^2$. In what follows, we focus on the case of stationary spatially localized patterns, so that we choose $B<1+A^2$ so that the system operates far from the Andronov-Hopf bifurcation for the stability of $(X_s,\,Y_s)$.

\begin{figure}[bbp]
\begin{center}
\includegraphics[width=1.0\textwidth]{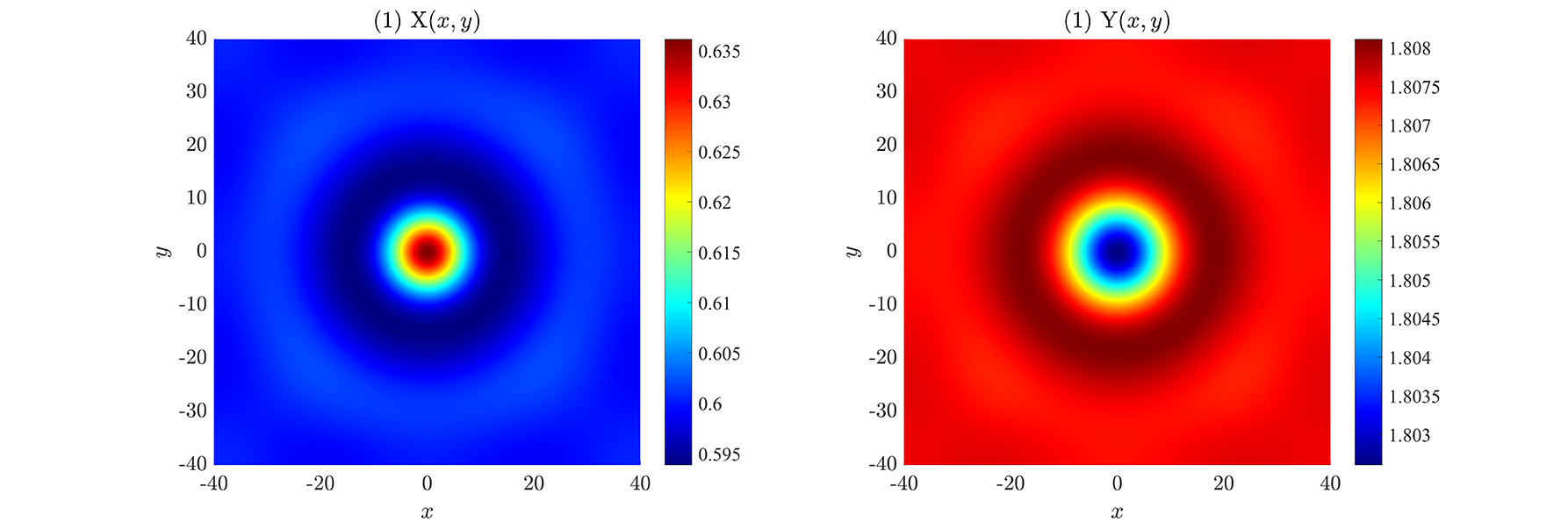}\\
\includegraphics[width=1.0\textwidth]{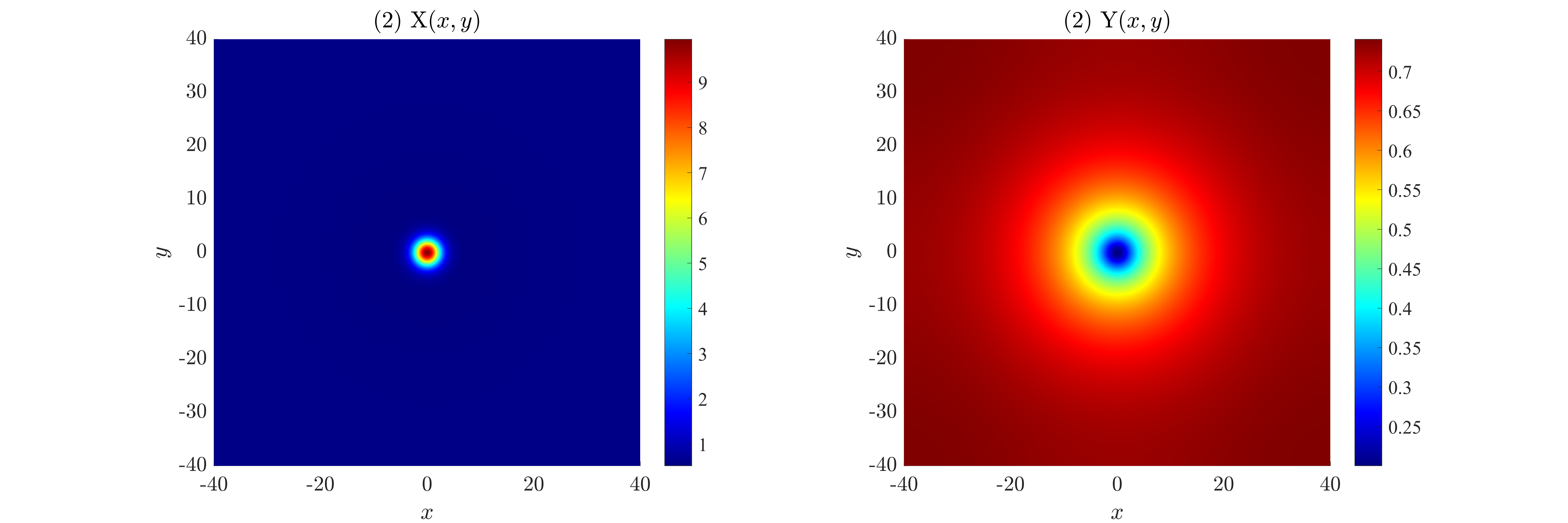}\\
\includegraphics[width=1.0\textwidth]{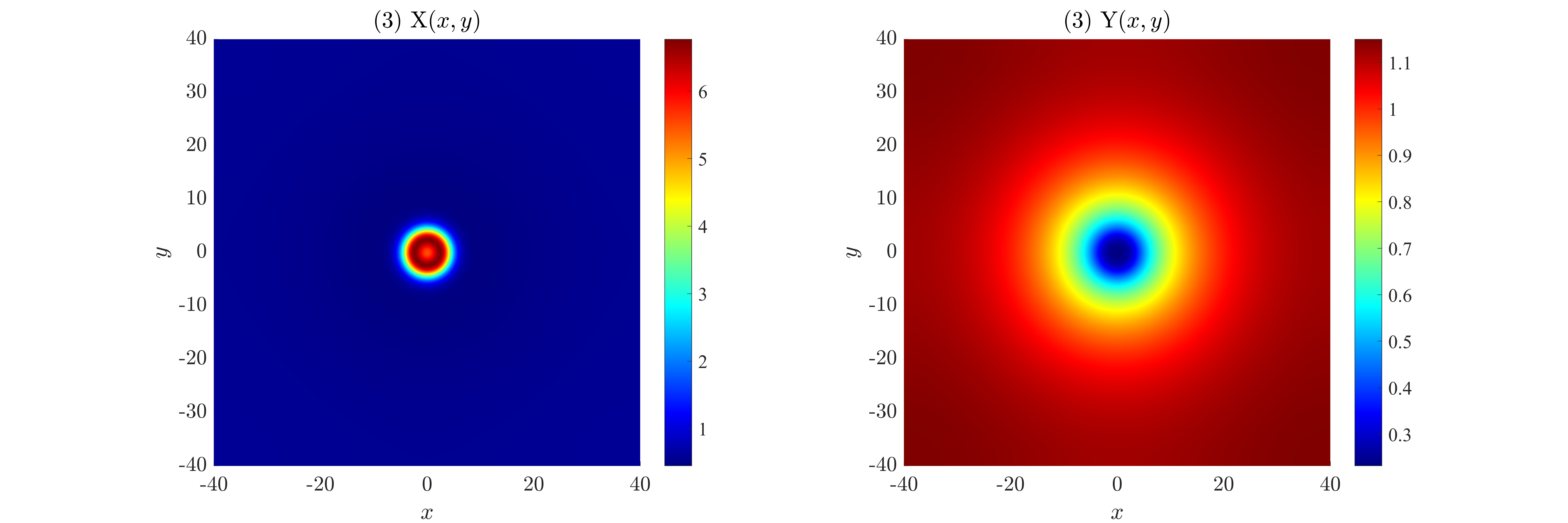}\\
\includegraphics[width=1.0\textwidth]{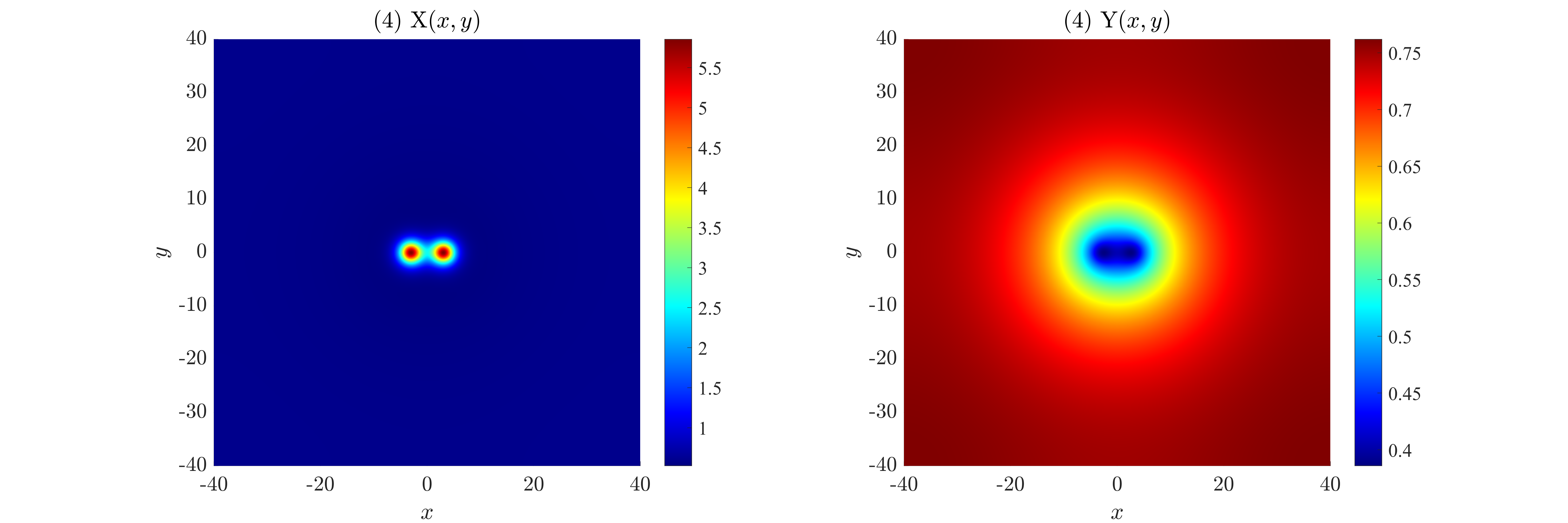}

\end{center}
\caption{ {Profiles of the single localized solution in the Brusselator model. First row: Unstable low-amplitude solution marked with (1) in Fig. \ref{bifdiag}. Second row: Stable localized solution marked with (2) in Fig. \ref{bifdiag}. Third row: Ringlike solution marked with (3) in Fig. \ref{bifdiag}. Fourth row: Self-replicated solution marked with (4) in Fig. \ref{bifdiag}. 
}}
\label{bifsol}
\end{figure}

The formation of spatially periodic structures such as hexagons or stripes and transitions between them have been intensively investigated for the  model~\eqref{eq:Bru} (see recent paper on this issue \cite{PenaPRE2001}). Besides periodic patterns, the same mechanism predicts the possible existence of aperiodic, localized
structures. They consist of isolated or randomly distributed  localized spots embedded in the uniform state \cite{RD_LS_7,Tlidi_Entropy,RD_LS_8}. Under uniform chemical constraints, non-equilibrium reaction-diffusion systems which exhibit a subcritical Turing-Prigogine instability  display  stable localized spot patterns in the pinning parameter range  where there is coexistence between two different states: a uniform branch of stationary state solutions and a branch of spatially periodic solutions.  In this hysteresis loop, a single circular spot consisting of a pinning front between a stable uniform solution and., e.g., a hexagonal branch of solutions is stable \cite{Tlidi_Entropy}. Indeed, the Brusselator model Eq. \eqref{eq:Bru}, supports stable two-dimensional localized solutions with periodic boundaries in both $x$ and $y$ directions 
as shown in Fig. \ref{bifsol} (2). One can observe these structures in direct numerical simulations of Eq.~\eqref{eq:Bru}, however this tool seems unfit for a systematic analysis of the bifurcational structure. We therefore deploy numerical continuation techniques provided by the matlab package pde2path \cite{pde2path}. Using this package we are able to track stable and unstable solutions as well as folds and other bifurcations. Providing this analysis in two spatial dimensions is a novelty in theoretical biology, however it has been successfully applied  in other fields of research such as e.g., modeling of thin films \cite{engel1,engel2} and nonlinear optics \cite{svetacontinuation}.

Starting from a stable localized solution obtained by direct numerical simulations with periodic boundaries, we explore the parameter space using the parameter $B$ as our primary continuation parameter. The $L_1$-norm of the activator field $X$, i.e.
\begin{equation}
 ||X||_{L_1}=\int|X-\overline{X}|\text{d}x\text{d}y\,
\end{equation} 
with $\overline{X}$ being the mean value of the field, is used as a measure to plot bifurcation diagrams.
Since the structures we are interested in possess a radial symmetry, we perform our calculations on a quarter of the real domain with Neumann-boundaries for the sake of numerical performance and then expand the plotted solutions shown in Fig. \ref{bifsol} and Fig. \ref{bifsol2}. For our calculations, we use a mesh adaption method both to refine the mesh where it is required and to coarse the mesh if possible. 
\begin{figure}[bbp]
	\begin{center}
		\includegraphics[width=1.0\textwidth]{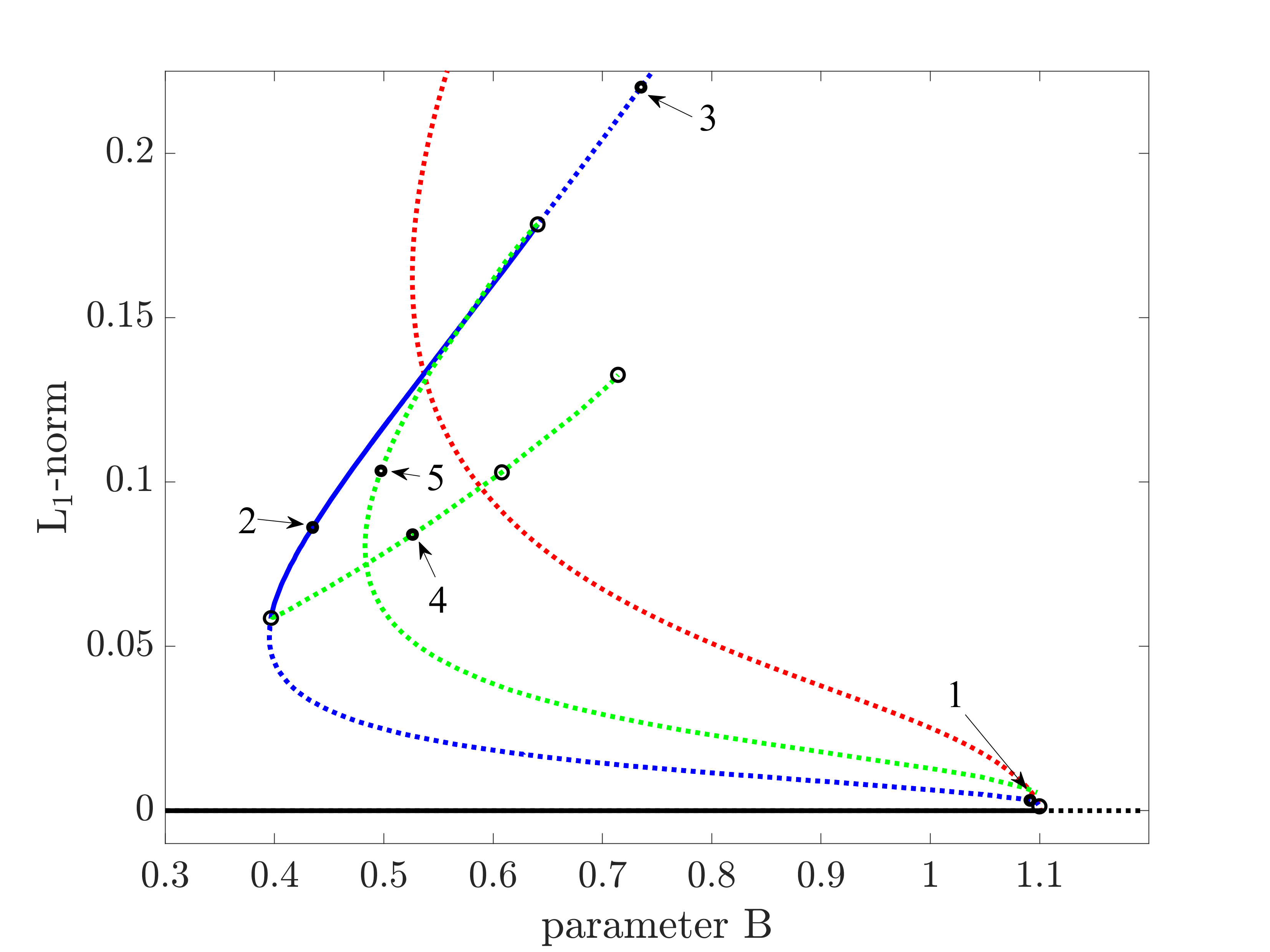}
	\end{center}
	\caption{{Bifurcation diagram for bound localized solutions in the two-dimensional Brusselator model \eqref{eq:Brudelay} without time-delayed feedback obtained similarly to Fig. \ref{bifdiag}. Shortly after the Turing-Prigogine point, the bound localized solution (dotted blue line) bifurcates from the periodic orbit in a supercritical pitchfork bifurcation. The localized solution becomes stable in a saddle-node bifurcation (solid blue line), where also an unstable solution branches off, with one of the structures decreasing in amplitude. The stable solution becomes then again unstable in a subcritical pitchfork bifurcation, in which a ringlike (dotted blue line) and a self-replicated solution (dotted green line) branch off. Solution profiles for the marked positions can be found in Fig. \ref{bifsol2}.}}
	\label{bifdiag2}
\end{figure}
\begin{figure}[bbp]
\begin{center}
\includegraphics[width=1.0\textwidth]{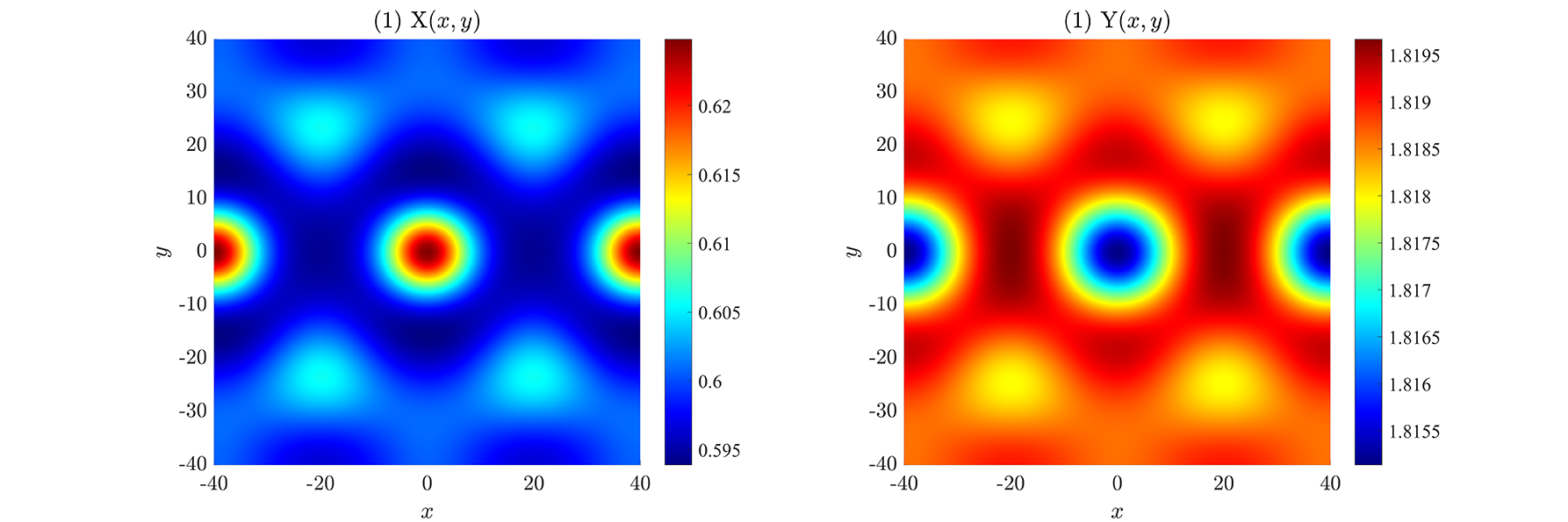}\\
\includegraphics[width=1.0\textwidth]{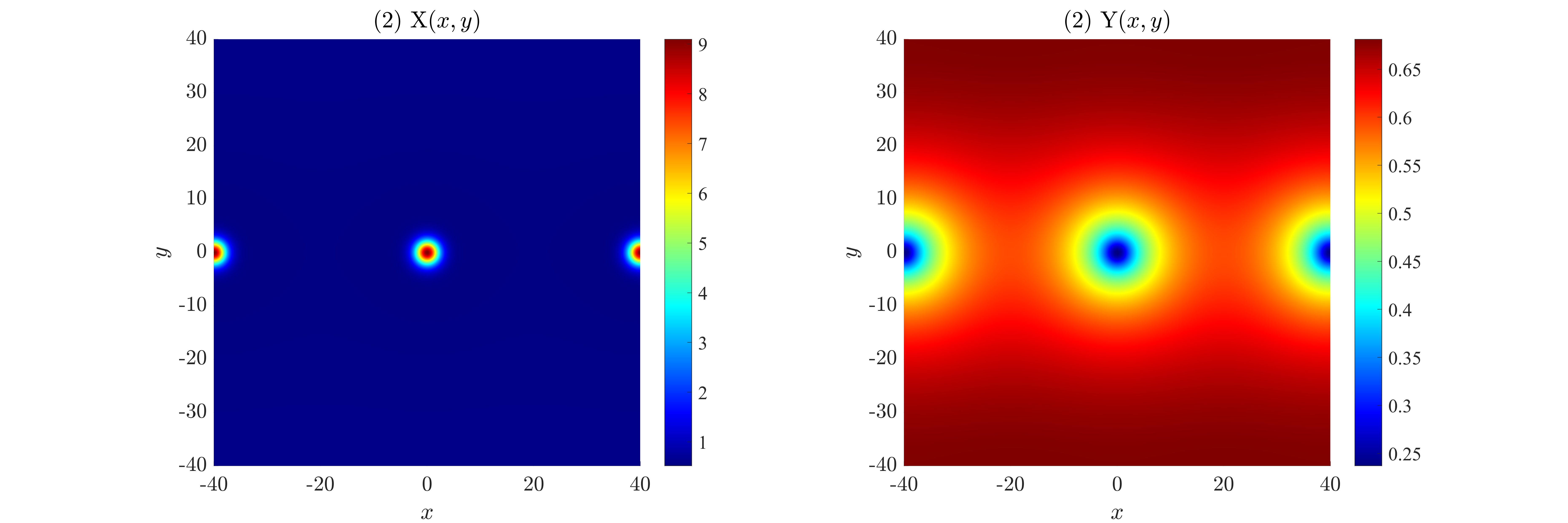}\\
\includegraphics[width=1.0\textwidth]{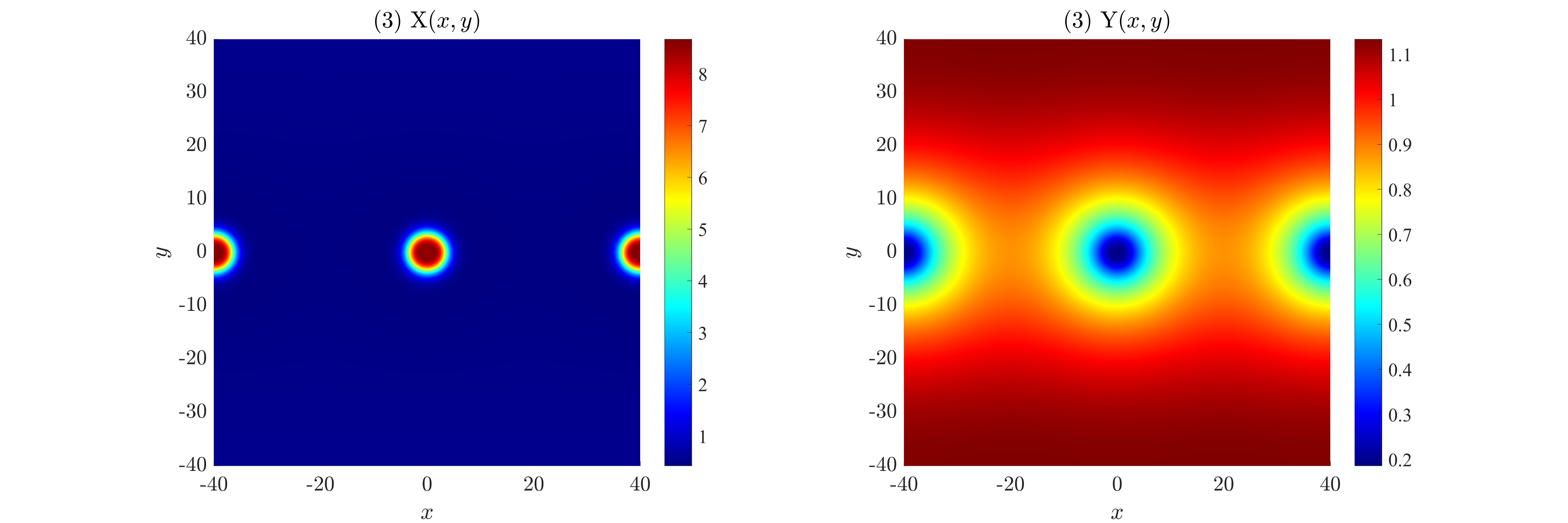}\\
\includegraphics[width=1.0\textwidth]{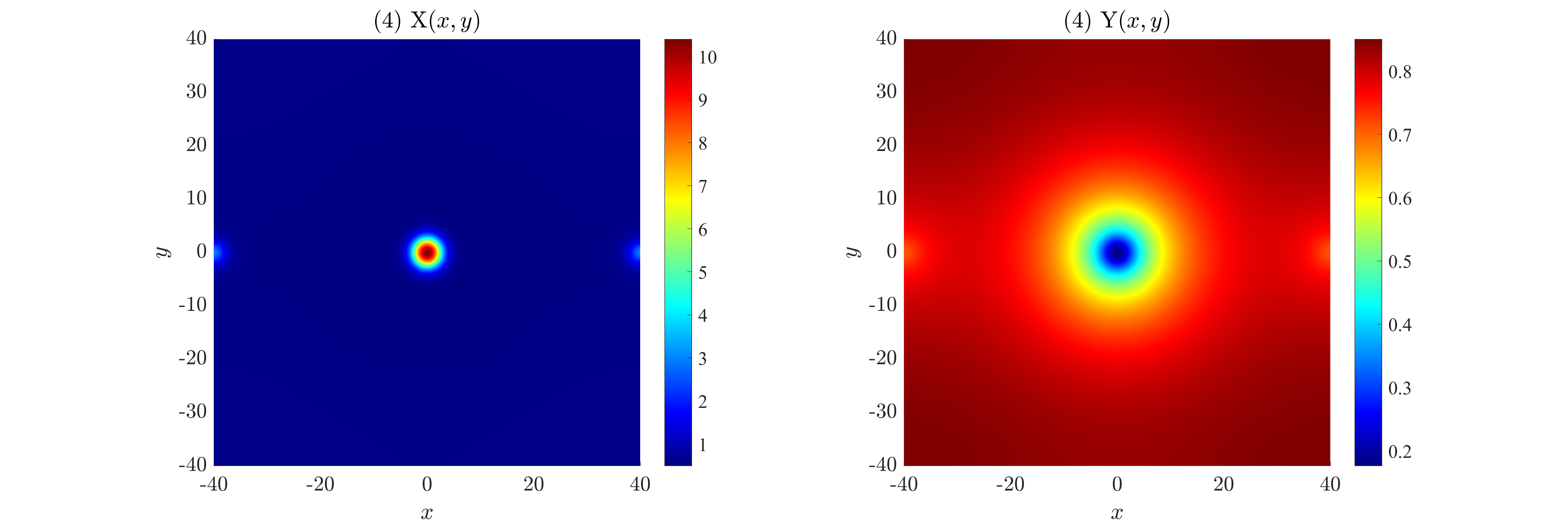}\\
\includegraphics[width=1.0\textwidth]{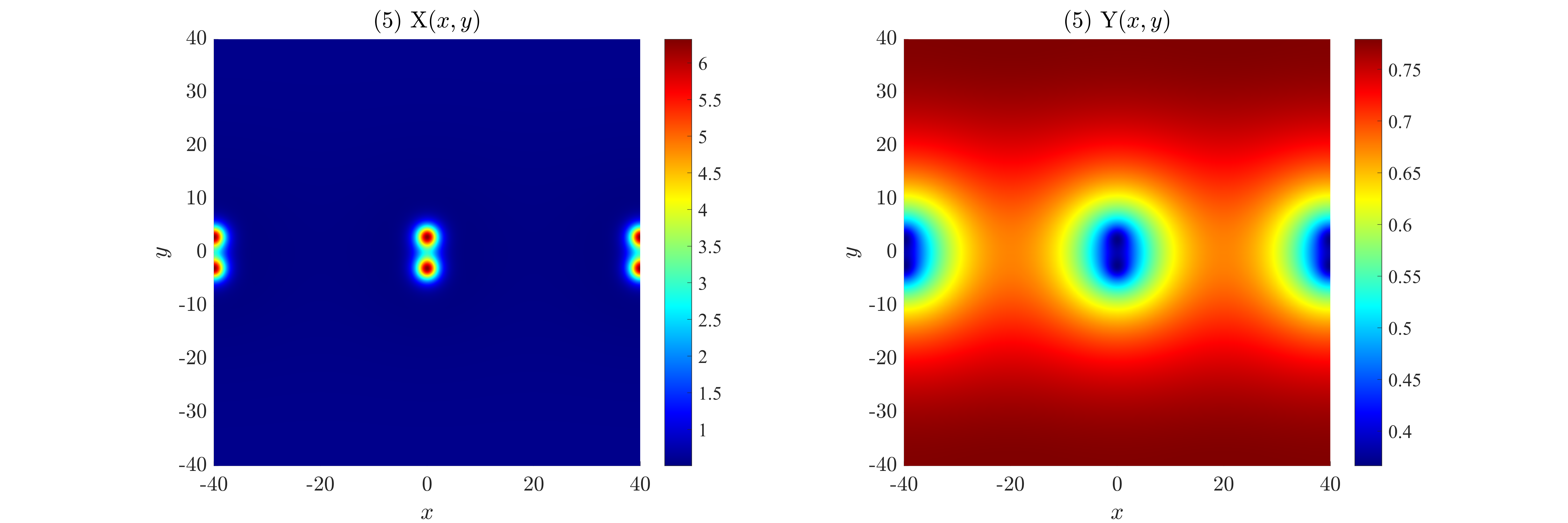}
\end{center}
\caption{ {Profiles of the bound localized solution in the Brusselator model. First row: Unstable low-amplitude solution marked with (1) in Fig. \ref{bifdiag2}. Second row: Stable localized solution marked with (2) in Fig. \ref{bifdiag2}. Third row: Ringlike solution marked with (3) in Fig. \ref{bifdiag2}. Fourth row: Unstable localized bound solution with one of the structures decreasing in amplitude marked with (4) in Fig. \ref{bifdiag2}. Fifth row: Self-replicated solution marked with (5) in Fig. \ref{bifdiag2}. 
}}
\label{bifsol2}
\end{figure}
\begin{figure}[bbp]
	\begin{center}
		\includegraphics[width=1.0\textwidth]{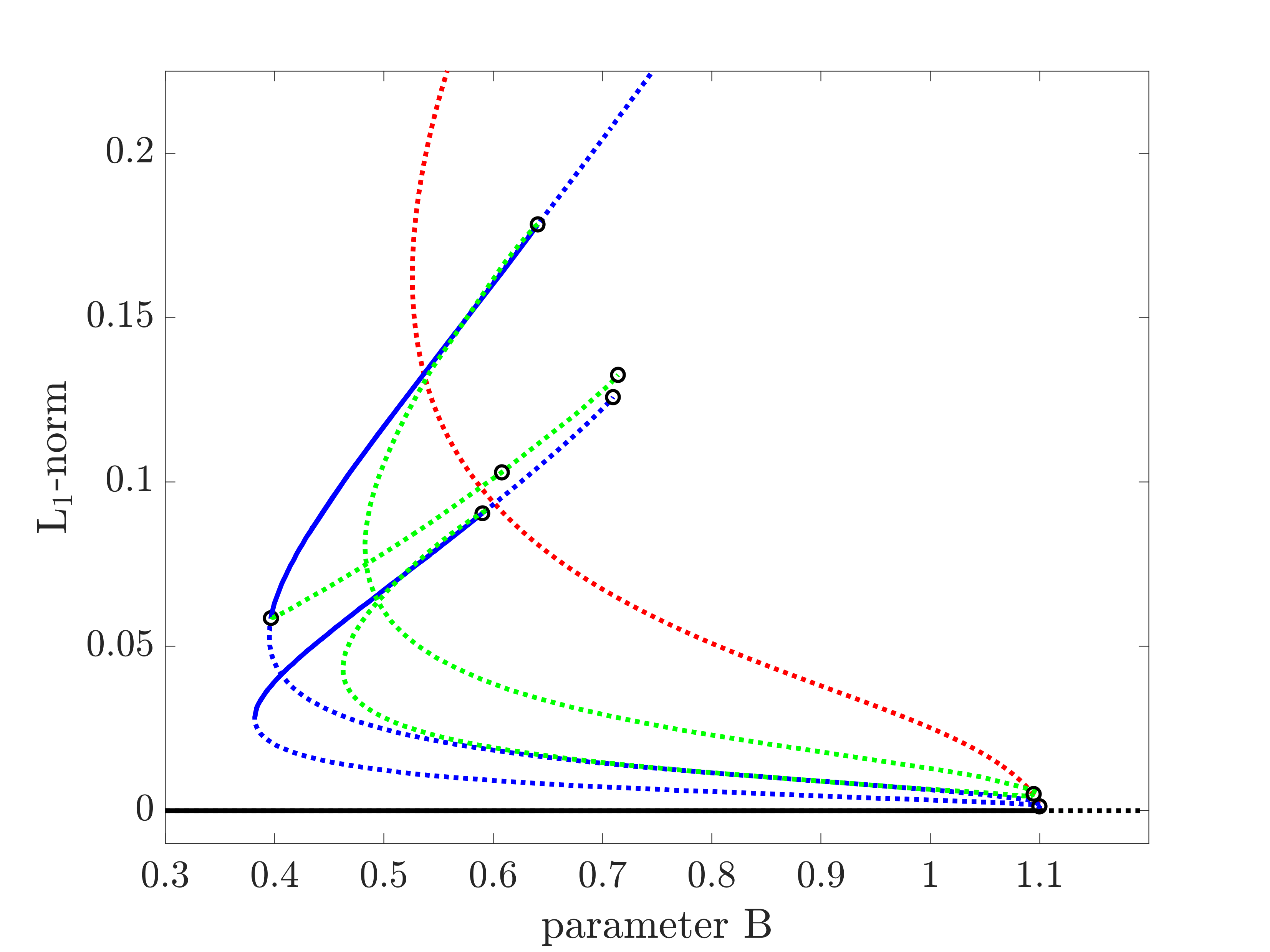}
	\end{center}
	\caption{{Bifurcation diagram for localized solutions in the two-dimensional Brusselator model \eqref{eq:Brudelay} without time-delayed feedback. Here the two localized solution branches (blue) depicted in Fig. \ref{bifdiag} and Fig. \ref{bifdiag2} are compared. The shape and the stability characteristics of both branches are very similar. The existence of other stable localized solutions is probable, but most of them will depend on the domain size alike the localized solution consisting of two spots (upper blue line). Furthermore, a green dotted solution branch exhibits qualitatively the same bifurcation scenario as the single localized spot solution does.}}
	\label{bifdiag3}
\end{figure}

The result of this bifurcation analysis is depicted in Fig. \ref{bifdiag}. The branch of localized solutions bifurcates from the branch of quadratic periodic solutions in a supercritical pitchfork bifurcation, very close to the critical Turing-point. Recent work on this topic suggests \cite{knobloch}, that this is a finite size effect and that in an infinite system, both branches would start at the critical point. At first, an unstable low amplitude solution starts to arise (cf. Fig. \ref{bifsol}, first row) that is, however, still unstable. Along the branch indicated by the blue dotted line in Fig. \ref{bifdiag} the structure grows in amplitude and finally reaches stability in a fold at $B\approx 0.4$. This is the main solution branch we are interested in, where a stable localized solution (cf. Fig. \ref{bifsol}, second row) exists approximately in the range of $0.4\le B \le 0.59$ (solid blue line). This regime of existence is in good agreement with results from direct numerical simulations. At $B\approx 0.59$, the localized solution loses its stability in a subcritical pitchfork bifurcation. Further along the now again unstable branch (blue dotted line), the structure develops a dip at the center, and transforms into a ring-like structure (cf. Fig. \ref{bifsol}, third row). It then runs through further bifurcations and becomes even more unstable, so we stop to trace the branch at $B\approx 0.71$. Also at the abovementioned bifurcation point, another unstable solution branches off (green dotted line). On this branch, the radial symmetry of the solution is broken. The solution shows a self-replicated structure of two bound localized structures in the first component $X$ of our system. In the faster diffusing second component $Y$ the solution looks more like an elliptic localized solution (cf. Fig. \ref{bifsol}, fourth row). This solution branch declines down to the periodic solution, suggesting, that it connects with a periodic branch, however, up to now we are not able to determine the type of bifurcation at the end of this branch.

Aside from the single localized solution, we were able to track another branch of localized solutions, also branching off from the periodic orbit in a similar way as the single solution. The bifurcation diagram is depicted in Fig. \ref{bifdiag2}. In this case, two equal localized spots form a bound solution with a distance of half the domain size between them. The solution starts again as an unstable low-amplitude solution (cf. Fig. \ref{bifsol2}, first row) and then reaches stability (solid blue line) through a fold and a nearby bifurcation, leading to the described bound state (cf. Fig. \ref{bifsol2}, second row). At this bifurcation, another solution branches off (dotted green line). Plotting this solution shows, that on this branch one of the two spots is decreasing (\ref{bifsol2}, fourth row). Therefore, the discrete translation symmetry of half the domain length is broken for solutions on this emerging branch. Similarly to the single spot solution, the bound solution loses its stability in a second bifurcation, and leads to a ringlike solution (\ref{bifsol2}, third row) which becomes more unstable through further bifurcations (not shown). At the second bifurcation point, another solution, where each spot is replicated branches off (dotted green line). The solution is depicted in Fig. \ref{bifsol2}, fifth row, and refers to the breaking of the radial symmetry. Although we were able to find this stable localized solution of two spots in direct numerical simulations also, one should note that this solution depends on the domain size. Contrary to the single localized solution, the exact shape changes with varying the domain size since, due to the periodic boundaries, every solution possesses two neighbors that interact with each other. 

All described solution branches are depicted in Fig. \ref{bifdiag3}.
One could suspect that the decreasing spot solution (Fig. \ref{bifsol2}, fourth row) branching off near the fold of the bounded localized solution connects to the single spot solution. But this is not the case since the decreasing spot does not vanish completely. Instead, it remains a bound solution of a smaller and a larger spot. One observes that the corresponding solution branch runs similarly through further bifurcations as the single localized spot solution does. However, in contrast to the single localized solution, it is unstable in the whole parameter space.

This section shows that, using numerical continuation techniques in two spatial dimensions, we are able to analyze the bifurcation structure of localized solutions. Most important for further investigations of localized solutions in the Brusselator model are the two regimes where stable localized solutions exist.  

\section{Branches of stationary solutions for the Brusselator model with delayed feedback}
In another line of research, it has been shown that time-delayed feedback can strongly impact the space-time dynamics of both periodic and localized states in various systems out of equilibrium described by  the Swift-Hohenberg equation with time-delayed feedback via the so-called Pyragas control \cite{Pyragas1992421, Tlidi_99,Panajotov_10,Averlant_12,Sveta_13,Svetlana_13,Tlidi_13}. In chemical reactions without diffusion, delay cannot be ignored in industrial applications
where recycling of unreacted reagents reduces the cost of the reaction. The output stream of a continuously stirred tank reactor  is sent
 through a separation process \cite{Erneux_book,Xia}. In reaction with diffusion and with global delayed feedback, a photoemission electron microscope was used to continuously image lateral distributions of adsorbed species, and  feedback is introduced by making the instantaneous dosing rate of the catalytic carbon monoxide oxidation-dependent on real-time properties of the imaged concentration patterns \cite{Mikhailov}.

The effect of spatial inhomogeneities for the delayed Swift-Hohenberg equation has been investigated both analytically and numerically in \cite{Tabbert} and a transition from oscillating to depinning solutions has been characterized. In this case, the time-delayed feedback acts as a driving force \cite{Tabbert}. The effect of noise responsible for the formation of dissipative structures in the delayed Swift-Hohenberg equation has been investigated recently in depth by Kuske et al. \cite{Kruske_17}. Furthermore, it was shown that delayed feedback can induce motion and breathing localized structures in chemical reaction-diffusion systems~\cite{Svetlana_13,Svetlana_PTRSA}. In advanced photonic devices the effect of the phase on the self-mobility of dissipative localized structures has been theoretically investigated in \cite{Pimenov, Schelte_C.:2017, Schemmelmann_T.:2017}. In driven Kerr cavities described by the Lugiato-Lefever equation, time-delayed feedback induces a drift of localized structures and the route to spatiotemporal chaos has been discussed in \cite{Panajotov_pra16,Akhmediev_16,Tlidi_chaos_17,Panajotov_epjd17}.

Recently, time-delayed feedback control has attracted a lot of interest in various fields of nonlinear science such as nonlinear optics, fiber optics, biology, ecology, fluid mechanics, granular matter, plant ecology (see recent overview \cite{Erneux_17}, and the excellent book by T. Erneux \cite{Erneux_book}. Numerical simulations of the Brusselator model with delayed feedback using Pyragas control have provided evidence of moving periodic structures in the form of stripes and hexagons \cite{Li_7} or superlattices \cite{Hu_08}. Dissipative structures and coherence resonance in the stochastic Swift-Hohenberg
equation with Pyragas control have also been investigated \cite{Kurske}.

We incorporate in the Brusselator model~\eqref{eq:Bru} a time-delayed feedback term.
The resulting evolution equations of the Brusselator model with time-delayed feedback read \cite{Tlidi_Entropy}
\begin{eqnarray}
\frac{\partial X}{\partial t} &= &\frac{\partial X^{2} }{\partial
	x^{2}} +\frac{\partial X^{2} }{\partial
	y^{2}}+ A-(B+1)X+X^2Y + \eta_x X(t-\tau)\nonumber\\
\frac{\partial Y}{\partial t} &= &D\left(\frac{\partial Y^{2} }{\partial x^{2}} +\frac{\partial Y^{2} }{\partial y^{2}}\right) + BX -X^2Y + \eta_y Y(t-\tau). \label{eq:Brudelay}
\end{eqnarray}
The delayed feedback parameters are the strengths of the feedbacks ($\eta_x,\,\eta_y$) and the delay time $\tau$. The approximation we use to model the delay term is valid in the limit where the strength of the delay is small. In this limit, the feedback concentrations are small enough for their evolution to be described by a single delay term. 

The homogeneous steady states with delayed feedback obey
\begin{eqnarray}
A &= & (B+1)X_s-\left[\frac{(1-\eta_x)X_s-A}{\eta_y}\right]X_s^2-\eta_xX_s\,,\nonumber\\
Y_s&= & \frac{(1-\eta_x)X_s-A}{\eta_y}\,.
\end{eqnarray}
The homogeneous steady states may be bistable. In what follows we focus our analysis on the monostable regime. The linear stability analysis of the homogeneous steady states with
respect to perturbations of the form $\exp{(i{\bf k}\cdot{\bf r} +\sigma t)}$ leads
to the transcendental characteristic equation
\begin{equation}
(\sigma+q^2+B+1-2X_sY_s-\eta_xe^{-\tau\sigma})(\sigma+Dq^2+X_s^2-\eta_ye^{-\tau\sigma})=X^2(B-2X_sY_s)\,.
\end{equation}

\begin{figure}
	\includegraphics[scale=0.57]{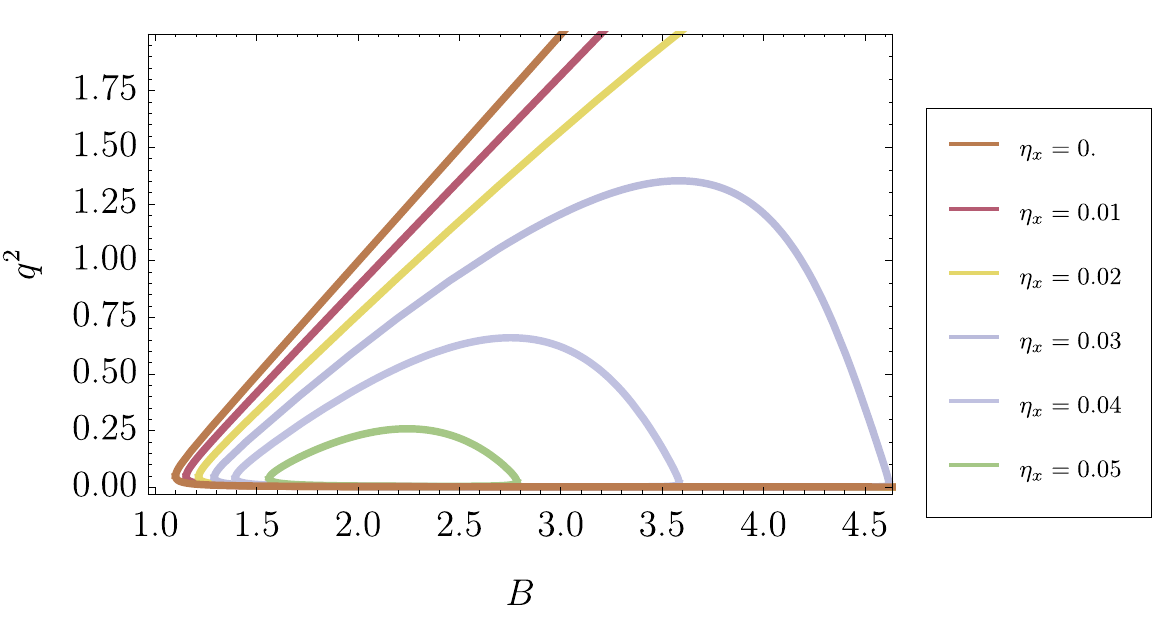}
	\includegraphics[scale=0.57]{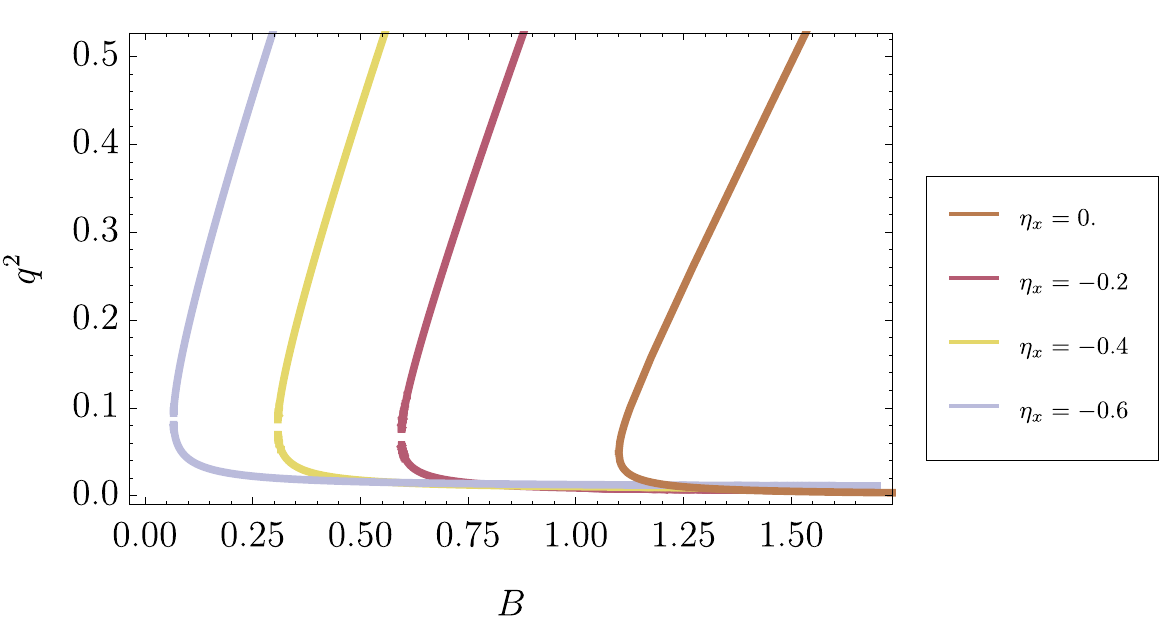}
	\caption{The evolution of the squared wavevector $q^2$ versus the control parameter $B$ calculated for the Turing-Prigogine instability. The feedback strength $\eta_x = -\eta_y$ is positive (left) and negative (right). Other parameters are: $A=0.4$, $D=150$.}
	\label{WavevectTuring}
\end{figure}

First, let us examine the case of a Turing-Prigogine instability where the eigenvalue $\sigma$ vanishes for a finite wavenumber $q=q_c$. Above the threshold associated with this instability, there exists a finite band of Fourier modes $q_-^2<q^2<q_+^2$ with
\begin{equation}
q_{\pm}^2=\frac{-F\pm\sqrt{F^2-4D[(1-\eta_x)X_s^2-\eta_y(B+1-2X_sY_s-\eta_x)]}}{2D},
\end{equation}
where $F=D(B+1-\eta_x-2X_sY_s)+X_s^2-\eta_y$. The Fourier modes are linearly unstable and trigger the spontaneous evolution of the state variable towards a stationary, spatially periodic distribution of the chemical concentration, which occupies the whole space available in the chemical open reactor. The critical wavenumber at the onset of the Turing-Prigogine instability reads
\begin{equation}
q_{c}^2=\frac{D(B+1-\eta_x-2X_sY_s)+X_s^2-\eta_y}{2D}
\end{equation}
and the threshold is a solution of the following equation
\begin{equation}
\left(D(B+1-\eta_x-2X_sY_s)+X_s^2-\eta_y\right)^2=4D[(1-\eta_x)X_s^2-\eta_y(B+1-2X_sY_s-\eta_x)]\,.
\end{equation}
The marginal curves associated with the Turing-Prigogine instability are shown in  Fig.~\ref{WavevectTuring} for different values of the feedback strength, $\eta_x=-\eta_y$ (the instability region is on the righthand side of the curves).
For negative values of $\eta_x$, the delay destabilizes the system and enables the Turing-Prigogine instability to occur for $B<B_{TP}=(1+A/D)^2$, where $B_{TP}$ is the threshold in the absence of delayed feedback. The threshold is thus shifted towards smaller $B$.
For positive values of $\eta_x$, the system restabilizes itself and only a finite domain of $B$ provides unstable Fourier modes. Increasing the delay strength further leads to the disappearance of the unstable band.\\

\begin{figure}[bbp]
\begin{center}
\includegraphics[scale=0.8]{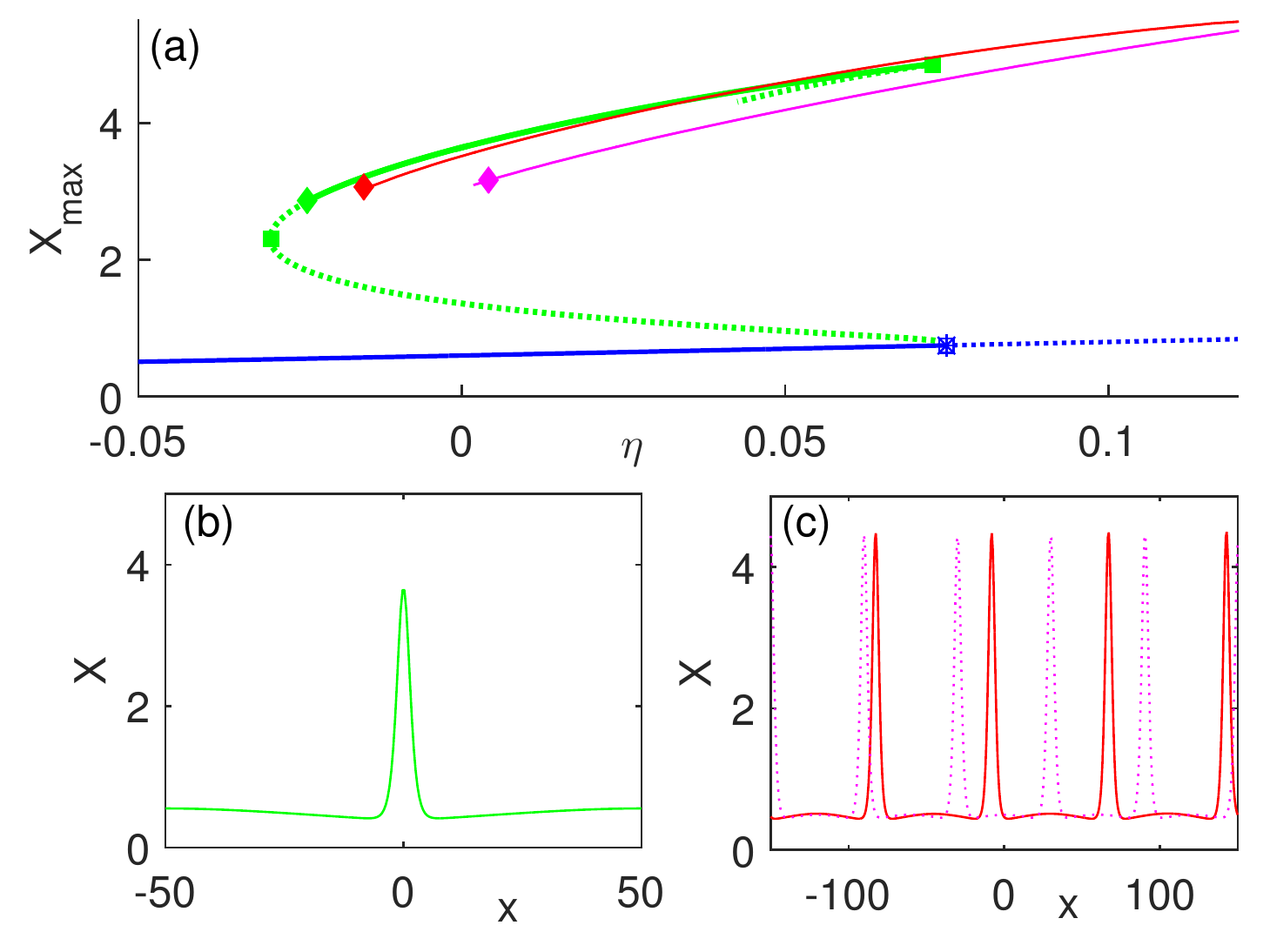}
\end{center}
\caption{ (a) One-dimensional bifurcation diagram of Eq.~\eqref{eq:Brudelay} in $(\eta,\,\mathrm{max}(X))$ plane ($\eta_x=\eta_y=\eta$) and fixed $\tau=10$. Green line: Localized solution branch for the $X$-component obtained by the Newton method with stable and unstable parts represented by solid and dashed line, respectively. This branch of localized solutions is delimited by an Andronov-Hopf bifurcation (diamond symbol) and a Saddle-Node (square symbol) bifurcations. Homogeneous steady state solution is represented by the blue line: solid and dashed for the stable and unstable case, respectively, delimited by a Turing bifurcation (star symbol). Red and magenta lines represent periodic solutions with different wavelengths. (b) and (c) show the spatial profiles in the selected window with periodic boundary conditions of the $X$ -component for the localized and periodic solutions with the same color as the corresponding branches presented in (a). The other parameters are: $A = 0.6$, $B = 0.8$ and $D = 150$.}
\label{Fig1}
\end{figure}

In Fig.~\ref{Fig1} (a) we present a one-dimensional bifurcation diagram for $\mathrm{max}(X)$ as a function of the feedback strength $\eta$ for $A=0.6$, $B=0.8$, $D=150$ and feedback parameters $\eta_x=\eta_y=\eta$ and $\tau=10$. Here, the green line represents the branch of localized solutions with a spatial profile as shown in Fig.~\ref{Fig1} (b). Numerical continuation of this branch, as a function of the feedback strength, has been carried out by the Newton method that allows finding both stable (represented by solid line) and unstable (represented by dashed line) states. The branch of localized solutions is limited by an Andronov-Hopf bifurcation (diamond symbol) and saddle-node (square symbol) bifurcations. The localized solution emerges from the homogeneous steady state represented by the blue line (solid and dashed for the stable and unstable parts, respectively), which is delimited by a subcritical Turing-Prigogine bifurcation (star symbol). Characteristic of this system is that it is highly multistable: not only a region of simultaneous stability of the localized and the homogeneous solution exists in Fig.~\ref{Fig1} (a), but there also exist stable branches of spatially periodic solutions with different wavelengths. Examples of two branches are shown in Fig.~\ref{Fig1} (a) represented by the red and the magenta lines. They are obtained by a direct numerical integration of Eqs.~\ref{eq:Brudelay}. The corresponding spatial profiles of $X$ are shown in Fig.~\ref{Fig1}~(c). Note that there exist several branches of moving LSs consisting of more peaks that are not shown in Fig. \ref{Fig1}.

\section{Transition from stationary to moving dissipative localized structures with delayed feedback}
\begin{figure}[h]
\begin{center}
\includegraphics[scale=0.5]{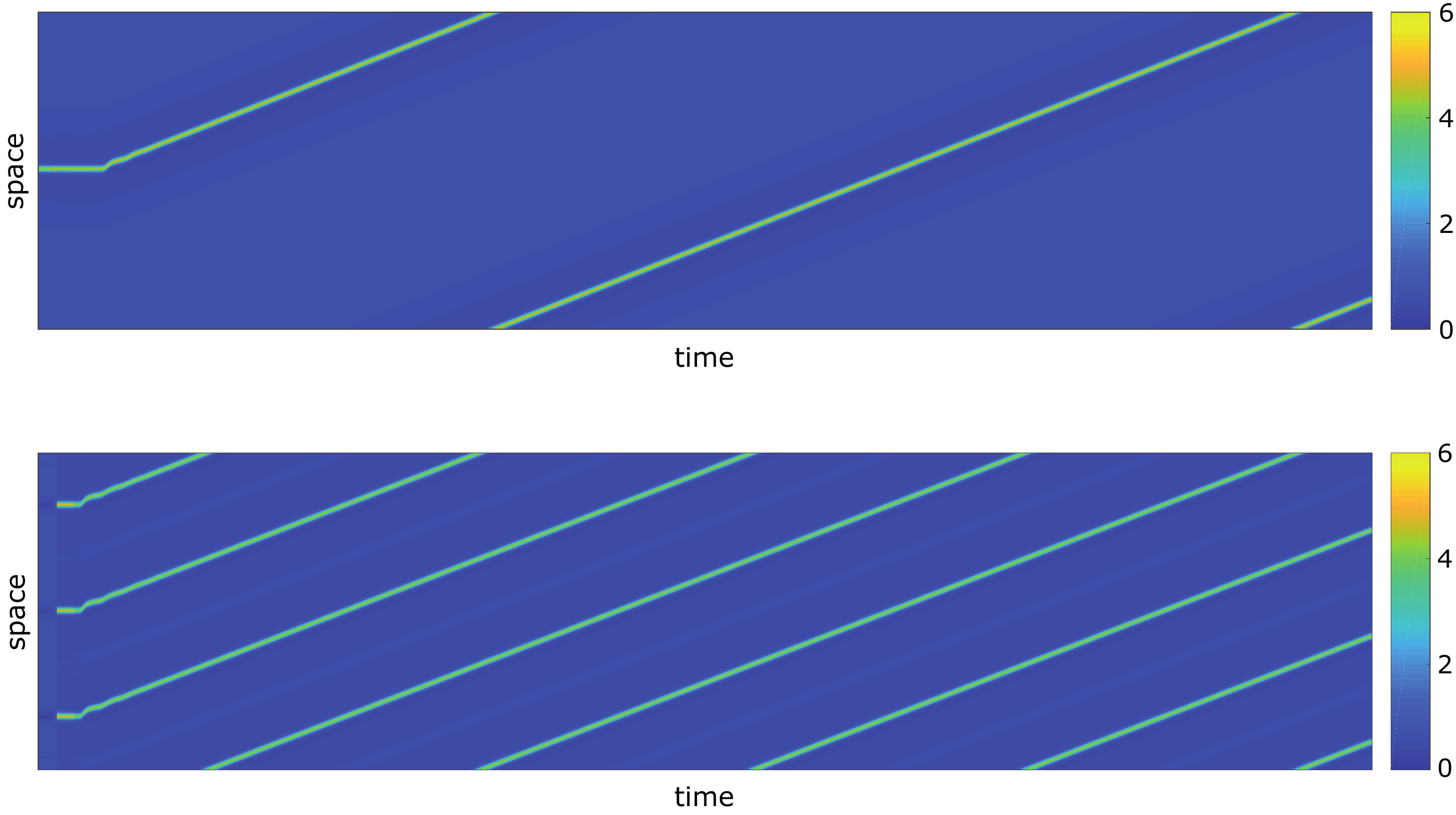}
\end{center}
\caption{Space-time representation of the $X$-component obtained by direct numerical simulation of the one-dimensional version of Eqs.~\eqref{eq:Brudelay} showing the time evolution from: (a) stable localised structure and (b) stable periodic pattern through moving structure and pattern, respectively. Parameters are $A = 0.6$, $B = 4.5$, and $D = 150$ and feedback with a fixed strength $\eta_x = - \eta_y = 0.05$ and delay time $\tau = 200$. The size of the system is a grid of $1024$x$1024$ with a spatial increment of dx $=0.1$.}
\label{Fig2}
\end{figure}

\begin{figure}[h]
\begin{center}
\includegraphics[scale=0.8]{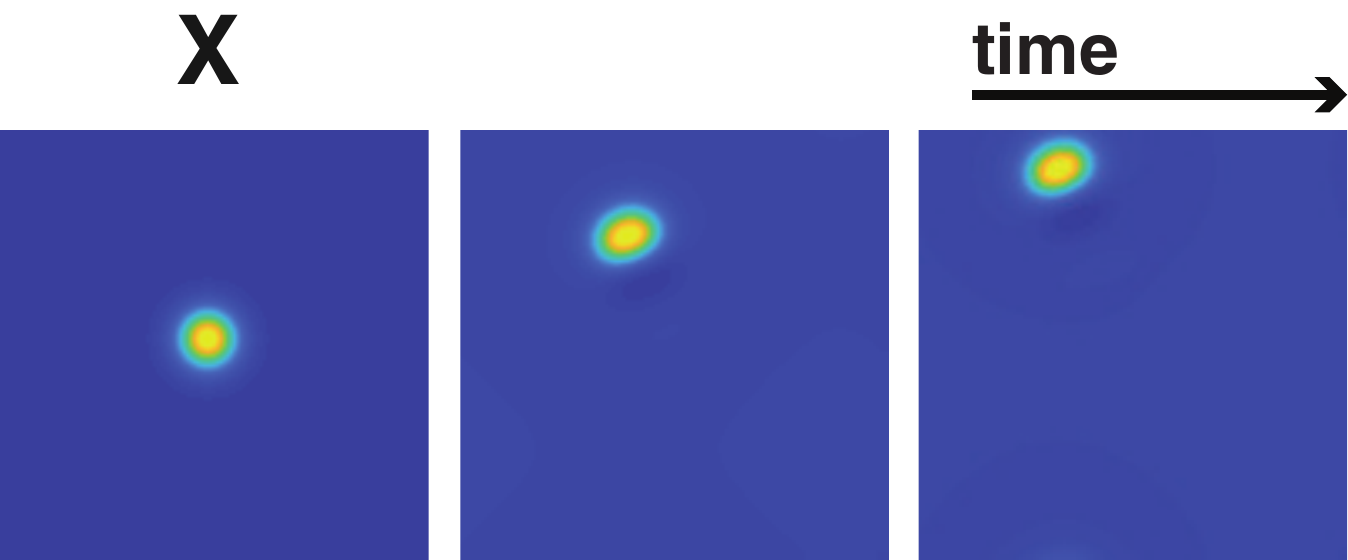}
\end{center}
\caption{A two-dimensional numerical simulation of Eqs.~\eqref{eq:Brudelay} showing a delay-induced drift of a single localized structure illustrated by snapshots in the $x-y$ plane of the $X$ variable at different times. Brusselator parameters are $A = 0.6$, $B = 0.45$  and $D = 150$ and feedback with a fixed strengths $\eta_x = - \eta_y = -0.17$ and delay time $\tau = 20$. The size of the system is a grid of $128$x$128$ with a spatial increment of dx $=0.3$.}
\label{Fig3}
\end{figure}

\begin{figure}[h]
\begin{center}
\includegraphics[scale=0.6]{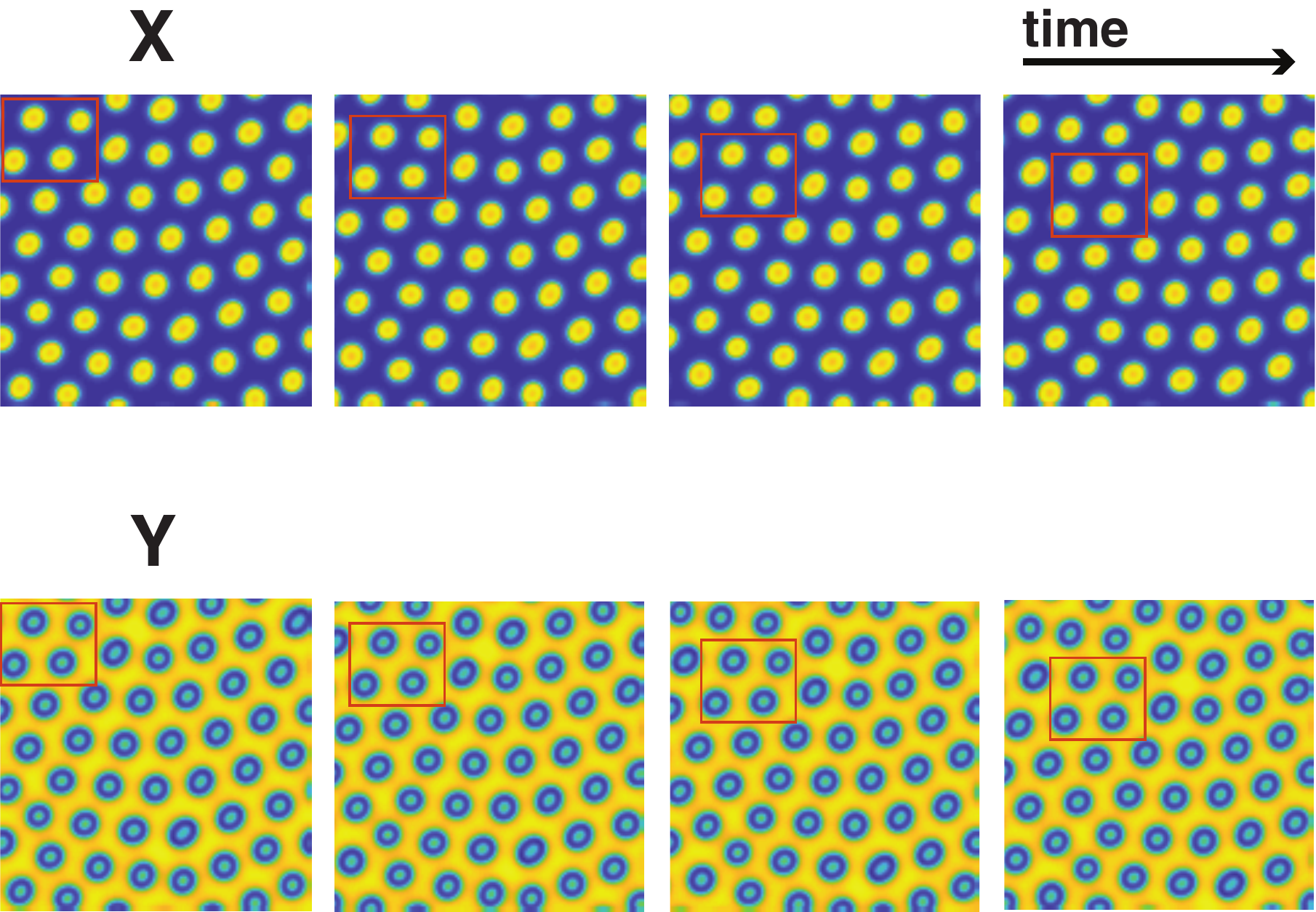}
\end{center}
\caption{A two-dimensional numerical simulation of Eqs.~\eqref{eq:Brudelay} showing a delay-induced drift of honeycomb pattern, found for Brusselator parameters of $A = 4.5$, $B = 30$ and $D = 8$ and feedback with a fixed strengths $\eta_x = - \eta_y = 0.1$ and delay time $\tau = 20$. The size of the system is a grid of $128$x$128$ with a spatial increment of dx $=0.2$. Red frames are added to help tracking the temporal evolution of the pattern.}
\label{Fi4}
\end{figure}

The delayed feedback may also induce a drift bifurcation of the localized structure in the Brusselator model, similar to that reported in the case of Swift-Hohenberg \cite{Tlidi_99}, Lugiato-Lefever \cite{Panajotov_epjd17} equations and the semiconductor laser model \cite{Panajotov_10,Pimenov,Schemmelmann_T.:2017,Schelte_C.:2017}. Such a feedback induced drift of a localized structure is illustrated in the one-dimensional case in Fig.~\ref{Fig2} (a) for the case of  $A = 0.6$, $B = 4.5$  and $D = 150$ and feedback with a fixed strength $\eta_x = - \eta_y = 0.05$ and delay time $\tau = 200$. We also illustrate the two-dimensional case. In Fig.~\ref{Fig3} we present an example of drifting localized structure for Brusselator parameters of $A = 0.6$, $B = 0.45$ and $D = 150$ and feedback with a fixed strength $\eta_x = - \eta_y = -0.17$ and delay time $\tau = 20$. We consider an initial condition which consists of stationary localized structures. During time evolution, while the structure remains quite symmetric and stationary (see Fig.~\ref{Fig3} (a)), it develops an asymmetric shape and starts moving along an arbitrary direction since the Brusselator model is isotropic in space as shown in the snapshots of Fig.~\ref{Fig3} (b) and (c). This drift then goes on indefinitely without change in speed or direction, as the LS goes through the boundary condition. Similarly, Fig.~\ref{Fi4} presents an example of drifting honeycomb pattern for Brusselator parameters of $A = 4.5$, $B = 30$ and $D = 8$ and feedback with a fixed strength $\eta_x = - \eta_y = -0.1$ while keeping the delay time the same as in the previous case. All numerical simulations are conducted for periodic boundary conditions. To initialize the simulation, the provided initial condition is also used as the feedback for the first $\tau$ units of time.

The localized structure is launched at $t=0$ and starts moving after a certain timespan, which decreases as the feedback strength is increased, see also \cite{Panajotov_10}. Furthermore, the localized structure moves faster as $\eta$ is increased. Our bifurcation analysis demonstrates that a localized state is a particular case since it coexists with stable periodic patterns with different wavelengths. Fig.~\ref{Fig2} (b) illustrates the case of delayed-induced drift for a periodic pattern. Note however that asymmetric LS solutions in 1D \cite{RD_LS_3,RD_LS_7} and in 2D \cite{Alejandro} have been reported even in the absence of the delayed feedback control.

In order to calculate the threshold of the delay-induced drift-bifurcation, we proceed in a similar way as in \cite{Tlidi_99,Pimenov,Panajotov_epjd17}. Slightly above this threshold we consider a localized structure moving uniformly with a constant small velocity $v=|{\mathbf v}|$ and expand the model variables $X$ and $Y$ in power series of $v$: $X = X_0(\boldsymbol{\xi}) + v \, \left[X_1(\boldsymbol{\xi}) + v X_2(\boldsymbol{\xi})+v^2 X_3(\boldsymbol{\xi})+...\right]$ and same expression for $Y$. Here $X_0(\boldsymbol{\xi}),\,Y_0(\boldsymbol{\xi})$ is the stationary localized structure profile, $\boldsymbol{\xi}=\mathbf{r}-v\,\mathbf{e}\,t$, $\mathbf{r}=(x,\,y)$, and $\mathbf{e}$ is the unit vector in the direction of motion of the localized structure. Substituting this expansion into Eq.~(\ref{eq:Brudelay}) and collecting the first order terms in the small parameter $v$, we obtain
\begin{eqnarray}
L \left(\begin{array}{l}X_{1}\\ Y_{1}\end{array}\right) = \left(\begin{array}{l}w_x(1 + \eta_x \tau)\\ w_y( 1 + \eta_y \tau) \end{array}\right)\label{eq:1storder}
\end{eqnarray}
with $w_x={\mathbf e}\cdot \nabla X_0$ and $w_y={\mathbf e}\cdot \nabla Y_0$. The linear operator $L$ is given by
$$
L = \left(\begin{array}{cc}\nabla^2-B-1+2X_0Y_0+\eta_x& X_0^2\\
B- 2X_0Y_0 &D\nabla^2-X_0^2+\eta_y
\end{array}\right).
$$
By applying the solvability condition to the right hand side of Eq.~(\ref{eq:1storder}), we get the drift instability threshold
\begin{equation}
\tau (\eta_x S_x + \eta_y S_y)= -(S_x+S_y)
\label{thresholdlimit}
\end{equation}
with $S_x=\langle \psi_x^{\dagger},\nabla X_0 \rangle$ and $S_y=\langle \psi_y^{\dagger},\nabla Y_0\rangle$. Here, the eigenfunction ${\psi}^{\dagger}=\left(\psi_x^{\dagger},\,\psi_y^{\dagger}\right)^T$  is the solution of the homogeneous adjoint problem $L^\dagger \psi^{\dagger} = 0$ and the scalar product $\langle \cdot , \cdot \rangle$ is defined as  $\langle U,\, V\rangle= \int_{-\infty}^{+\infty}U V\,d{\mathbf r}$.
For feedback strengths of $\eta_x=\eta_y=\eta$, we recover from (\ref{thresholdlimit}) the threshold condition obtained earlier for the drift instability of cavity solitons in the Swift-Hohenberg equation with delayed feedback, $\eta\tau = -1$ \cite{Tlidi_99}. 

One can also calculate the drift threshold for the case of $-\eta_x=\eta_y=\eta$, where we obtain:
\begin{equation}
\tau\eta = \frac{\langle \psi_x^{\dagger},\nabla X_0 \rangle+\langle \psi_y^{\dagger},\nabla Y_0\rangle}{\langle \psi_x^{\dagger},\nabla X_0 \rangle-\langle \psi_y^{\dagger},\nabla Y_0\rangle} = \frac{1+\chi}{1-\chi},
\label{thresholdlimit2}
\end{equation}

with $\chi=\langle \psi_y^{\dagger},\nabla Y_0\rangle / \langle \psi_x^{\dagger},\nabla X_0 \rangle$. In the one-dimensional case, for the parameters $A = 0.6$, $B = 0.8$ and $D = 150$, with a delay time of $\tau = 20$ this expression gives an estimation for the feedback strength of $\eta_{analytical} = 0.049 = \eta_{numerical}$, which corresponds perfectly to the numerically estimated threshold. In the two-dimensional case, for the same parameters as Fig.~\ref{Fig3}, we reach the same conclusion with $\eta_{analytical} = 0.050$.

\begin{figure}	
	\centering
	\includegraphics[scale=0.75]{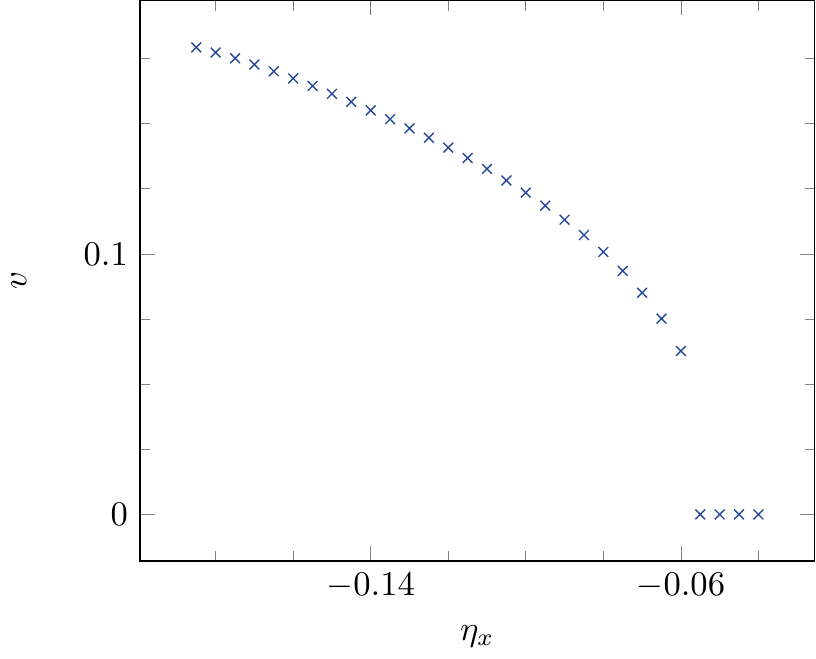}
	\caption{Velocity of the moving localized structure $v$ versus the feedback strength $\eta_x =-\eta_y$ obtained by numerical simulation. Parameters are the same as in Fig.~\ref{Fig3}.}
	\label{speeds}
\end{figure}

One can also calculate the velocity of the structure close to the bifurcation point. The derivation is very similar to the one performed \cite{Svetlana_13} for the case of Pyragas control \cite{Pyragas1992421}, so it will not be performed in full detail here. We choose the Ansatz
\begin{align}
(X,Y)^{T}=\textbf{q}(\textbf{r},t)=\textbf{q}_0(\textbf{r}-\textbf{R}(t)),
\end{align}
where, $\textbf{q}$ is a vector containing both components $X$ and $Y$, $\textbf{q}_0$ is the stationary solution, $\textbf{r}=(x,y)^{T}$ is a vector containing the spatial coordinates and $\textbf{R}(t)$ is the position of the localized solution. Inserting this Ansatz into Eq. \eqref{eq:Brudelay} and projecting on the adjoint eigenfunction $\boldsymbol{\psi}^{\dagger}=(\psi_x^{\dagger},\psi_y^{\dagger})^{T}$ yields:

\begin{align}
\dot{\textbf{R}}=-\eta\left\{\left[\textbf{R}(t)-\textbf{R}(t-\tau)\right]-\frac{\beta}{6}\left[\textbf{R}(t)-\textbf{R}(t-\tau)\right]^3\right\}
\end{align}
with
\begin{align}
\beta=\frac{\langle\boldsymbol\nabla\boldsymbol\psi^\dagger|\boldsymbol\nabla\boldsymbol\psi\rangle}{\langle\boldsymbol\psi^\dagger|\boldsymbol\psi\rangle}.
\end{align}
Expanding $\textbf{R}(t-\tau)$ leads to the evolution equation for the velocity:
\begin{align}
\frac{\tau^2\eta}{2}\dot{\textbf{v}}=(\eta\tau+1)\textbf{v}-\frac{\beta}{6}\eta\tau^3\textbf{v}^3,\label{eq.vel}
\end{align}
showing the normal form of a supercritical pitchfork bifurcation with the bifurcation at $\eta\tau=-1$. In accordance to the results mentioned above, at $\eta\tau=-1$ Eq. \eqref{eq.vel} changes from the trivial stable solution $\textbf{v}=\boldsymbol{0}$, to a nontrivial solution with
\begin{align}
|\textbf{v}|=\pm\frac{1}{\tau}\sqrt{\frac{\eta\tau+1}{\beta\eta\tau}},
\end{align}
i.e. the solution starts to drift. Note that this approximation is only valid close to the bifurcation point since any shape deformations due to additional stable modes were neglected in the Ansatz.\\
This pitchfork bifurcation is confirmed by numerical simulation (see Fig.~\ref{speeds}).

\section*{Acknowledgment}

Stimulating discussions with Thomas Erneux are gratefully acknowledged. 
M.T. received support from the Fonds National de la Recherche Scientifique (Belgium). K.P. acknowledges the Methusalem Foundation. F.T. received funds from the German scholarship foundation and the Center for Nonlinear Science M{\"u}nster.
 
\section{Conclusions}
We have investigated the formation of dissipative localized structures in the paradigmatic Brusselator model in a two-dimensional setting. Using path-continuation techniques, we have characterized the transition from a single spot towards periodic patterns through a self-replication phenomenon. The branch of localized solutions bifurcates from the quadratic periodic solutions in a supercritical pitchfork bifurcation, very close to the critical Turing-Prigogine point. The localized solution loses its stability in a subcritical pitchfork bifurcation and develops a dip at the center, transforming it into a ring-like structure. At the pitchfork bifurcation point another unstable solution, with broken radial symmetry, branches off. The solution shows a self-replicated structure of two bound localized structures. We have also found another branch of two localized spots with a distance of half the domain size between them. Similarly to the single spot solution, this bound solution loses its stability in second bifurcations, leading to a ringlike solution on one branch and to replicated spots on the other branch. 

In addition, we have investigated the influence of the delay feedback control on the dynamics of localized and periodic structures in the Brusselator model. In particular, we have shown that delay feedback can induce the motion of one and two-dimensional localized spots in an arbitrary direction since the system is isotropic in the $(x,y)$ direction. When increasing further the strength of the feedback and keeping the delay time constant, the steady state chemical concentration exhibits a complex dynamical behavior induced by the delayed feedback. In particular, it would be interesting to characterize the spatiotemporal chaos and the formation of rogue waves in two-dimensional settings. We shall come back to these questions in a separate publication.


\begin{thebibliography}{9}

\bibitem{Balesu}Prigogine I,  Balescu R. 1956 Ph\'enom\`enes cycliques dans la thermodynamique des processus irr\'eversibles,  Bull. Cl. Sci. Acad. R. Belg. XLII, 256--265.

\bibitem{Lefever}Prigogine I, Lefever R. 1968 Symmetry breaking instabilities in dissipative systems. II,  \textit{J. Chem. Phys}. \textbf{48}, 1695--1700. (\href{https://doi.org/10.1063/1.1668896}{doi:10.1063/1.1668896})

\bibitem{Prigogine}Glansdorff P, Prigogine I. 1971 \textit{Thermodynamic theory of structures, stability and fluctuations}. Wiley New York.

\bibitem{Lefever_PTRSA}Lefever R. 2018 The rehabilitation of irreversible processes and dissipative structures' 50th
anniversary. \textit{Phil. Trans. R. Soc. A} \textbf{376}, 20170365. (\href{https://doi.org/10.1098/rsta.2017.0365}{doi:10.1098/rsta.2017.0365})

\bibitem{ERneux_PTRS}Erneux T. 2018 Early models of chemical oscillations failed to provide bounded solutions. \textit{Phil.
Trans. R. Soc. A} \textbf{376}, 20170380. (\href{https://doi.org/10.1098/rsta.2017.0380}{doi:10.1098/rsta.2017.0380})

\bibitem{Turing}Turing AM. 1952 The chemical basis of morphogenesis. \textit{Phil. Trans. R. Soc. Lond. B} {\bf 237}, 37--72. (\href{https://doi.org/10.1098/rstb.1952.0012}{doi:10.1098/rstb.1952.0012})

\bibitem{DeKepper}Castets V, Dulos E, Boissonade J, and De Kepper P. 1990 Experimental evidence of a sustained standing Turing-type nonequilibrium chemical pattern.
\textit{Phys. Rev. Lett}, \textbf{64}, 2953--2956. (\href{http://dx.doi.org/doi:10.1103/PhysRevLett.64.2953}{doi:10.1103/PhysRevLett.64.2953})

\bibitem{Swinney} Ouyang Q, Swinney HL. 1991 Transition from a uniform state to hexagonal and striped Turing patterns. \textit{Nature}, \textbf{352}, 610--612. (\href{http://dx.doi.org/doi:10.1038/352610a0}{doi:10.1038/352610a0})

\bibitem{Rev5}Rosanov NN. 2002 \textit{Spatial Hysteresis and Optical Patterns}. Springer, Berlin. (\href{https://doi.org/10.1007/978-3-662-04792-7}{doi:10.1007/978-3-662-04792-7})

\bibitem{Rev6}Staliunas K, Sanchez-Morcillo VJ. 2003 \textit{Transverse Patterns in Nonlinear Optical Resonators}. Springer Tracts in Modern Physics Vol. 188. Springer-Verlag, Berlin Heidelberg. (\href{https://doi.org/10.1007/3-540-36416-1}{doi:10.1007/3-540-36416-1})

\bibitem{Rev9}Mandel P, Tlidi M. 2004 Transverse dynamics in cavity nonlinear optics (2000-2003). \textit{J. Opt. B:
Quant. Semiclass. Opt.}. \textbf{6}, R60. (\href{https://doi.org/10.1088/1464-4266/6/9/R02}{doi:10.1088/1464-4266/6/9/R02})


\bibitem{Rev11}Mikhailov AS,  Showalter K. 2006 Control of waves, patterns and turbulence in chemical systems. Phys. Rep. \textbf{425}, 79. (\href{https://doi.org/10.1016/j.physrep.2005.11.003}{doi:10.1016/j.physrep.2005.11.003})



\bibitem{Rev12}Tlidi M, Kolokolnikov T, Taki M. 2007 Introduction: dissipative localized structures in extended systems. \textit{Chaos}, \textbf{17}, 037101. (\href{https://doi.org/10.1063/1.2786709}{doi:10.1063/1.2786709})

\bibitem{Rev13}Akhmediev N, Ankiewicz A. 2008 \textit{Dissipative Solitons: From Optics to Biology and Medicine}. Springer-Verlag, Berlin, Heidelberg. (\href{https://doi.org/10.1007/978-3-540-78217-9}{doi:10.1007/978-3-540-78217-9})

\bibitem{Rev16}Purwins HG, Bodeker HU, Amiranashvili S. 2010  Dissipative solitons. \textit{Advances in Physics}, \textbf{59}, 485-701. (\href{https://doi.org/10.1080/00018732.2010.498228}{doi:10.1080/00018732.2010.498228})


\bibitem{Lugiatobook} Lugiato L, Prati F, Brambilla M. 2015 \textit{Nonlinear Optical Systems}. Cambridge University Press. (\href{https://doi.org/10.1017/CBO9781107477254}{doi:10.1017/CBO9781107477254})

\bibitem{Tlidi-Krassi}Tlidi M, Panajotov K. 2018 Cavity solitons: Dissipative structures in Nonlinear Photonics, \textit{Romanian Reports in Physics}, \textbf{70}, 406.

\bibitem{Rev8}Murray JD. 2003  \textit{Mathematical Biology}, 3de ed. (Springer, Berlin). (\href{https://doi.org/10.1007/B98868}{doi:10.1007/B98868})

\bibitem{Deneubourg}Deneubourg JL, Goss S, Franks N, Sendova-Franks A, Detrain C, Chr\'etien L. 1991  The dynamics of collective sorting robot-like ants and ant-like robots. \textit{Proceedings of the first international conference on simulation of adaptive behavior on From animals to animats} (pp. 356-363). MIT press.

\bibitem{ga96} Goldbeter A. 1996 \textit{Biochemical Oscillations and Cellular Rhythms.} CUP, Cambridge.


\bibitem{kURAMOTO-kOGA}Koga S, Kuramoto Y. 1980 Localized Patterns in Reaction-Diffusion Systems. {\em Progress of Theoretical Physics}, \textbf{63}, 106. (\href{https://doi.org/10.1143/PTP.63.106}{doi:10.1143/PTP.63.106})


\bibitem{Malomed_Ne}Malomed BA, Nepomnyashchy AA. 1990 Kinks and solitons in the generalized Ginzburg-Landau equation. {\em Phys. Rev. A}, \textbf{42}, 6009--6014. (\href{https://doi.org/10.1103/PhysRevA.42.6009}{doi:10.1103/PhysRevA.42.6009})

\bibitem{Tlidi_94}Tlidi M, Mandel P, Lefever R. 1994 Localized structures and localized patterns in optical bistability. {\em Physical Review Letters}, \textbf{73}, 640. (\href{https://doi.org/10.1103/PhysRevLett.73.640}{doi:10.1103/PhysRevLett.73.640})
  
  
\bibitem{Dewel_94}Jensen O, Pannbacker VO, Mosekilde E, Dewel G, Borckmans P. 1994 Localized structures and front propagation in the Lengyel-Epstein model.  {\em Physical Review E}, \textbf{50}, 736-749. (\href{http://dx.doi.org/10.1103/PhysRevE.50.736}{doi:10.1103/PhysRevE.50.736})

\bibitem{Hilali}Hilali MF, M{\'e}tens S, Borckmans P, Dewel G. 1995 Pattern selection in the generalized Swift-Hohenberg model, {\em Physical Review E}, \textbf{51}, 2046. (\href{https://doi.org/10.1103/PhysRevE.51.2046}{doi:10.1103/PhysRevE.51.2046})

\bibitem{Leblond-Mihalache} Leblond H, Mihalache D. 2013 Models of few optical cycle solitons beyond the slowly varying envelope approximation. {\em Physics Reports}, \textbf{523}, 61--126. (\href{https://doi.org/10.1016/j.physrep.2012.10.006}{doi:10.1016/j.physrep.2012.10.006})


\bibitem{Tlidi-PTRA}Tlidi M, Staliunas K, Panajotov K, Vladimirov AG, Clerc M. 2014 Introduction: Localized structures in dissipative media: from optics to plant ecology. {\em Phil. Trans. R. Soc. A}, \textbf{372}, 20140101. (\href{https://doi.org/10.1098/rsta.2014.0101}{doi:10.1098/rsta.2014.0101})

\bibitem{RD_LS_1}Doelman A, Eckhaus W, Kaper TJ. 2000 Slowly-modulated two-pulse solutions in the Gray-Scott
model I: asymptotic construction and stability, SIAM J. Appl. Math. \textbf{61}, 1080-1102. (\href{https://doi.org/10.1137/S0036139999354923}{doi:10.1137/S0036139999354923})

\bibitem{RD_LS_2}Doelman A, Eckhaus W, Kaper TJ. 2001 Slowly-modulated two-pulse solutions in the Gray-Scott
model II: geometric theory, bifurcations and splitting dynamics, SIAM J. Appl. Math., \textbf{61}, 2036-2062. (\href{https://doi.org/10.1137/S0036139900372429}{doi:10.1137/S0036139900372429})

 \bibitem{RD_LS_3}Ward MJ, Wei JC. 2002 The existence and stability of asymmetric spike patterns in the Schnakenberg
model, Studies in Appl. Math., \textbf{109}, 229-264. (\href{https://doi.org/10.1111/1467-9590.00223}{doi:10.1111/1467-9590.00223})


\bibitem{RD_LS_4}Kolokolnikov T, Ward MJ, Wei JC. 2005 The existence and stability of spike equilibria in the
one-dimensional Gray-Scott model: The pulse-splitting regime, Physica D: Nonlinear Phenomena, \textbf{202}, 258-293. (\href{http://dx.doi.org/10.1016/j.physd.2005.02.009}{doi:10.1016/j.physd.2005.02.009})


\bibitem{RD_LS_5}Kolokolnikov T, Tlidi M. 2007 Spot deformation and replication in the
  two-dimensional belousov-zhabotinski reaction in a water-in-oil
  microemulsion, Phys. Rev. Lett. \textbf{98}, 188303. (\href{https://doi.org/10.1103/PhysRevLett.98.188303}{doi:10.1103/PhysRevLett.98.188303})


\bibitem{RD_LS_6}Kolokolnikov T, Ward MJ, Wei JC. 2009 Spot self-replication and dynamics for the Schnakenberg model in a two-dimensional domain, J. Nonlinear Science, \textbf{19}, 1-56. (\href{http://doi.org/10.1007/s00332-008-9024-z}{doi:10.1007/s00332-008-9024-z})


\bibitem{RD_LS_7}Tzou JC, Bayliss A, Matkowsky BJ, Volpert VA. 2011 Stationary and slowly moving localized
pulses in a singularly perturbed Brusselator model, European J. Appl. Math., \textbf{22}, 423-453.(\href{https://doi.org/10.1017/S0956792511000179}{doi:10.1017/S0956792511000179})


\bibitem{Tlidi_Entropy}Tlidi M, Gandica Y, Sonnino G, Averlant E, Panajotov K. 2016 Self-Replicating spots in the brusselator model and extreme events in the one-dimensional case with delay, \textit{Entropy}, \textbf{18}, 64. (\href{http://dx.doi.org/10.3390/e18030064}{doi:10.3390/e18030064})

\bibitem{RD_LS_8}Tzou JC, Ward MJ. 2018 The stability and slow dynamics of spot patterns in the 2D Brusselator
model: The effect of open systems and heterogeneities, Physica D: Nonlinear Phenomena, \textbf{373}, 13-37. (\href{https://doi.org/10.1016/j.physd.2018.02.002}{doi:10.1016/j.physd.2018.02.002})

\bibitem{TVL_11}Vladimirov AG, Lefever R, Tlidi M. 2011 Relative stability of multipeak localized patterns of cavity solitons. {\em Physical Review A}, \textbf{84}, 043848. (\href{http://dx.doi.org/10.1103/PhysRevA.84.043848}{doi:10.1103/PhysRevA.84.043848})


\bibitem{Lier2013} Liehr A. 2013 {\em Dissipative Solitons in Reaction Diffusion Systems. Mechanism,
	Dynamics, Interaction}, volume~70 of {\em Springer Series in Synergetics}. Springer Berlin/Heidelberg. (\href{https://doi.org/10.1007/978-3-642-31251-9}{doi:10.1007/978-3-642-31251-9})


\bibitem{Pearson} Pearson JE. 1993 Complex patterns in a simple system. {\em Science}, \textbf{261}, 189--192. (\href{https://doi.org/10.1126/science.261.5118.189}{doi:10.1126/science.261.5118.189})


\bibitem{LEE} Lee K, McCormick WD, Pearson JE, Swinney HL. 1994 Experimental observation of self-replicating spots in a reaction–diffusion system. {\em Nature}, \textbf{369}, 215--218. (\href{https://doi.org/10.1038/369215a0}{doi:10.1038/369215a0})

\bibitem{pde2path}
Uecker H, Wetzel D, Rademacher J. 2014 pde2path - a matlab package for continuation and bifurcation in 2d
elliptic systems. {\em Numerical Mathematics: Theory, Methods and Applications},
\textbf{7}:58--106. (\href{https://doi.org/10.1017/S1004897900000295}{doi:10.1017/S1004897900000295})

\bibitem{PenaPRE2001}Pe\~na B, P\'erez C. 2001 Stability of Turing patterns in the Brusselator model,
{\em Phys. Rev. E}, \textbf{64}, 056213. (\href{https://doi.org/10.1103/PhysRevE.64.056213}{doi:10.1103/PhysRevE.64.056213})

\bibitem{engel1}Engelnkemper S, Wilczek M, Gurevich S.V, Thiele U. Morphological transitions of sliding drops: Dynamics and bifurcations,  {\em Phys. Rev. Fluids}, \textbf{1}, 073901. (\href{https://doi.org/10.1103/PhysRevFluids.1.073901}{doi:10.1103/PhysRevFluids.1.073901})

\bibitem{engel2}Engelnkemper S, Gurevich SV, Uecker H, Wetzel D, Thiele U. Continuation for thin film hydrodynamics and related scalar problems, \textit{Computational Modeling of Bifurcations and Instabilities in Fluid Mechanics}, Springer, 2018. 

\bibitem{svetacontinuation}Schelte C, Javaloyes J, Gurevich SV. 2018 Dynamics of temporally localized states in passively mode-locked semiconductor lasers, {\em Phys. Rev. A}, \textbf{97}, 053829. (\href{https://doi.org/10.1103/PhysRevA.97.053820}{doi:10.1103/PhysRevA.97.053820})

\bibitem{knobloch}Bergeon A, Burke J, Knobloch E, Mercader I. 2008 Eckhaus instability and homoclinic snaking,
{\em Phys. Rev. E}, \textbf{78}, 046201. (\href{https://doi.org/10.1103/PhysRevE.78.046201}{10.1103/PhysRevE.78.046201})


\bibitem{Tlidi_99}Tlidi M, Vladimirov AG, Pieroux D, Turaev, D. 2009 Spontaneous motion of cavity solitons induced by a delayed feedback. {\em Phys. Rev. Lett.}, {\bf 103}, 103904. (\href{https://doi.org/10.1103/PhysRevLett.103.103904}{doi:10.1103/PhysRevLett.103.103904})




\bibitem{Panajotov_10}Panajotov K, Tlidi M. 2010 Spontaneous motion of cavity solitons in vertical-cavity lasers subject to optical injection and to delayed feedback. {\em Eur. Phys. J. D}, {\bf 59}, 67--72. (\href{https://doi.org/10.1140/epjd/e2010-00111-y}{doi:10.1140/epjd/e2010-00111-y})


\bibitem{Averlant_12}Tlidi M, Averlant E, Vladimirov A, Panajotov K. 2012 Delay feedback induces a spontaneous motion of two-dimensional cavity solitons in driven semiconductor microcavities. {\em Phys.
Rev. A}, {\bf 86}, 033822. (\href{https://doi.org/10.1103/PhysRevA.86.033822}{doi:10.1103/PhysRevA.86.033822}) 

\bibitem{Tlidi_13}Tlidi M, Sonnino A, Sonnino G. 2013 Delayed feedback induces motion of localized spots in reaction-diffusion systems. {\em Phys. Rev. E}, {\bf 87}, 042918. (\href{https://doi.org/10.1103/PhysRevE.87.042918}{doi:10.1103/PhysRevE.87.042918}) 

\bibitem{Sveta_13}Gurevich SV, Friedrich R. 2013 Instabilities of localized structures in dissipative systems with delayed feedback. Physical Review Letters. {\bf 110}, 014101. (\href{https://doi.org/10.1103/PhysRevLett.110.014101}{doi:10.1103/PhysRevLett.110.014101})

\bibitem{Svetlana_13}Gurevich SV. 2013 Dynamics of localized structures in reaction-diffusion systems induced by delayed feedback. {\em Phys. Rev. E}, {\bf 87}, 052922. (\href{https://doi.org/10.1103/PhysRevE.87.052922}{doi:10.1103/PhysRevE.87.052922}) 


\bibitem{Pyragas1992421}
Pyragas K. 1992 Continuous control of chaos by self-controlling feedback. {\em Phys. Lett. A}, \textbf{170}(6):421--428. (\href{https://doi.org/10.1016/0375-9601(92)90745-8}{doi:10.1016/0375-9601(92)90745-8}) 

\bibitem{Erneux_book}Erneux T. 2009 \textit{Applied delay differential equations}, 3, Springer Science. (\href{https://doi.org/10.1007/978-0-387-74372-1}{doi:10.1007/978-0-387-74372-1})

\bibitem{Xia} Xia S, Hodge NY, Wiesner TF. 2009 Closed-form solution to the problem of reaction with fixed bed adsorption using delay-differential equations.  Chemical Engineering Science, Volume \textbf{64}, Issue 9. 2057-2066 (\href{https://doi.org/10.1016/j.ces.2009.01.025}{doi:10.1016/j.ces.2009.01.025})

\bibitem{Mikhailov}Kim M, Bertram M, Pollmann M, von Oertzen A, Mikhailov AS, Rotermund HH,  Ertl G. 2001 Controlling chemical turbulence by global delayed feedback: pattern formation in catalytic CO oxidation on Pt (110). {\em Science},  {\bf 292}, 1357-1360. (\href{https://doi.org/10.1126/science.1059478}{doi:10.1126/science.1059478})

\bibitem{Tabbert}Tabbert F, Schelte C, Tlidi M, Gurevich SV. 2017 Delay-induced depinning of localized structures in a spatially inhomogeneous Swift-Hohenberg model. {\em Phys. Rev. E}, {\bf 95}, 032213.  (\href{https://doi.org/10.1103/PhysRevE.95.032213}{doi:10.1103/PhysRevE.95.032213}) 

\bibitem{Kruske_17}Kuske R, Lee CY,Rottsch{\"a}fer V. 2018 Patterns and coherence resonance in the stochastic Swift-Hohenberg equation with Pyragas control: The Turing bifurcation case. {\em Physica D: Nonlinear Phenomena}, {\bf 365}, 57--71.
(\href{https://doi.org/10.1016/j.physd.2017.10.012}{doi:10.1016/j.physd.2017.10.012}) 

\bibitem{Svetlana_PTRSA}Gurevich SV. 2014 Time-delayed feedback control of breathing localized structures in a three-component reaction-diffusion system. {\em Phil. Trans. R. Soc. A}, \textbf{372}, 20140101. (\href{https://doi.org/10.1098/rsta.2014.0014}{doi:10.1098/rsta.2014.0014})


\bibitem{Pimenov}Pimenov A, Vladimirov AG, Gurevich SV, Panajotov K, Huyet G, Tlidi M. 2013 Delayed feedback control of self-mobile cavity solitons. {\em Phys. Rev. A}, \textbf{88}, 053830. (\href{https://doi.org/10.1103/PhysRevA.88.053830}{doi:10.1103/PhysRevA.88.053830})

\bibitem{Schelte_C.:2017}
Schelte C, Panajotov K, Tlidi M, Gurevich SV. 2017 Bifurcation structure of cavity soliton dynamics in a vertical-cavity surface-emitting laser with a saturable absorber and time-delayed feedback. {\em Phys. Rev. A}, \textbf{96}:023807. (\href{https://doi.org/10.1103/PhysRevA.96.023807}{doi:10.1103/PhysRevA.96.023807})

\bibitem{Schemmelmann_T.:2017}
Schemmelmann T, Tabbert F, Pimenov A, Vladimirov AG, Gurevich SV. 2017 Delayed feedback control of self-mobile cavity solitons in a wide-aperture laser with a saturable absorber. {\em Chaos: An Interdisciplinary Journal of Nonlinear Science},
\textbf{27}:114304. (\href{https://doi.org/10.1063/1.5006742}{doi:10.1063/1.5006742})

\bibitem{Panajotov_pra16}Panajotov K, Puzyrev D, Vladimirov AG, Gurevich SV, Tlidi M. 2016 Impact of time-delayed feedback on spatiotemporal dynamics in the Lugiato-Lefever model. {\em Phys. Rev. A}, \textbf{93}, 043835. (\href{https://doi.org/10.1103/PhysRevA.93.043835}{doi:10.1103/PhysRevA.93.043835})

\bibitem{Tlidi_chaos_17}Tlidi M, Panajotov K, Ferr\'e M, Clerc MG. 2017 Drifting cavity solitons and dissipative rogue waves induced by time-delayed feedback in Kerr optical frequency comb and in all fiber cavities. {\em Chaos: An Interdisciplinary Journal of Nonlinear Science}. {\bf 27}, 114312. (\href{https://doi.org/10.1063/1.5007868}{doi:10.1063/1.5007868})


\bibitem{Akhmediev_16}Akhmediev N et al. 2016 Roadmap on optical rogue waves and extreme events. Journal of Optics. {\bf 18}, 063001. (\href{https://doi.org/10.1088/2040-8978/18/6/063001}{doi:10.1088/2040-8978/18/6/063001})

\bibitem{Panajotov_epjd17}Panajotov K, Clerc MG, Tlidi M. 2017 Spatiotemporal chaos and two-dimensional dissipative rogue waves in Lugiato-Lefever model. {\em Eur. Phys. J. D}, \textbf{71}, 176. (\href{https://doi.org/10.1140/epjd/e2017-80068-y}{doi:10.1140/epjd/e2017-80068-y})

\bibitem{Erneux_17}Erneux T, Javaloyes J, Wolfrum M, Yanchuk, S. 2017 Introduction to Focus Issue: Time-delay dynamics. {\em Chaos: An Interdisciplinary Journal of Nonlinear Science}, \textbf{27}, 114201. (\href{https://doi.org/10.1063/1.5011354}{doi:10.1063/1.5011354})


\bibitem{Li_7}Li QS, Hu HX. 2007 Pattern transitions induced by delay feedback. {\em The Journal of chemical physics}, \textbf{127}, 00154510. (\href{https://doi.org/10.1063/1.2792877}{doi:10.1063/1.2792877})


 \bibitem{Hu_08}Hu HX, Li QS, Ji L. 2008 Superlattice patterns and spatial instability induced by delay feedback. {\em Physical Chemistry Chemical Physics}, \textbf{10}, 438. (\href{https://doi.org/10.1039/b712567d}{doi:10.1039/b712567d})

 \bibitem{Kurske}Kuske R, Lee CY, Rottschafer V. 2017 Patterns and coherence resonance in the stochastic Swift-Hohenberg equation with Pyragas control: The Turing bifurcation case. Physica D: Nonlinear Phenomena, \textbf{365}, 57-71. (\href{https://doi.org/10.1016/j.physd.2017.10.012}{doi:10.1016/j.physd.2017.10.012})


\bibitem{Alejandro} Alvarez-Socorro AJ, Clerc MG, Tlidi M. 2018 Spontaneous motion of localized structures induced by parity symmetry breaking transition. Chaos: An Interdisciplinary Journal of Nonlinear Science. \textbf{28}, 053119. (\href{https://doi.org/10.1063/1.5019734}{doi:10.1063/1.5019734})



\end{thebibliography}
\end{document}